\documentclass[11pt]{article}

\usepackage[final]{acl}

\usepackage{times}
\usepackage{latexsym}
\usepackage{pifont}
\usepackage{amsmath}
\usepackage{amssymb}
\usepackage{fvextra}
\usepackage{amsthm}

\usepackage{booktabs}
\usepackage{tabularx}
\usepackage{multirow}
\usepackage{float}
\usepackage{xcolor}
\input{borland.style.minted}
\DefineVerbatimEnvironment{MintedVerbatim}{Verbatim}{%
    breaklines=true,%
    breakanywhere=true,%
    fontsize=\scriptsize,%
    frame=lines,%
    framesep=2mm%
}
\newcommand{\setminted}[1]{}
\newfloat{listing}{tbp}{lop}
\floatname{listing}{Listing}
\usepackage{xcolor}

\usepackage{enumitem} 
\usepackage{amssymb} 

\usepackage[T1]{fontenc}

\usepackage[utf8]{inputenc}

\usepackage{microtype}

\usepackage{inconsolata}

\usepackage{graphicx}
\usepackage{array}
\usepackage{multirow}
\definecolor{brightgreen}{HTML}{00A82D}

\usepackage{xcolor}
\usepackage[most]{tcolorbox} 
\newcommand\blfootnote[1]{%
  \begingroup
  \renewcommand\thefootnote{}\footnote{#1}%
  \addtocounter{footnote}{-1}%
  \endgroup
}
\newtcolorbox{promptbox}[1][]{
  colback=gray!5!white,    
  colframe=gray!75!black,  
  title={\textbf{Prompt}}, 
  fonttitle=\bfseries,     
  breakable,               
  enhanced,                
  boxrule=1pt,           
  sharp corners=south,     
  #1                       
}

\title{SWE-Mutation: Can LLMs Generate Reliable Test Suites in Software Engineering?}

\author{
 \textbf{Yuxuan Sun\textsuperscript{1}},
 \textbf{Yuze Zhao\textsuperscript{1}},
 \textbf{Yufeng Wang\textsuperscript{2}},
 \textbf{Yao Du\textsuperscript{3}},
 \textbf{Zhiyuan Ma\textsuperscript{1}},
 \\
 \textbf{Jinbo Wang\textsuperscript{4}},
 \textbf{Mengdi Zhang\textsuperscript{5}},
 \textbf{Kai Zhang\textsuperscript{1}},
 \textbf{Zhenya Huang\textsuperscript{1,6}\thanks{Corresponding author}},
\\
 \textsuperscript{1}State Key Laboratory of Cognitive Intelligence, University of Science and Technology of China
 \\
 \textsuperscript{2}Independent Researcher.
 \textsuperscript{3}Beihang University.
 \textsuperscript{4}School of Mathematical Sciences, Peking University
 \\
 \textsuperscript{5}NeoShell AI.
 \textsuperscript{6}Institute of Artificial Intelligence, Hefei Comprehensive National Science Center
\\
   \small\texttt{\{sunyuxuan, yuzezhao, zhyma\}@mail.ustc.edu.cn, \{yfwang1118, mdzhangmd\}@gmail.com}\\
    \small\texttt{duyaoo@buaa.edu.cn, wangjinbo@stu.pku.edu.cn, \{kkzhang08, huangzhy\}@ustc.edu.cn}
 }
\begin{document}
\maketitle
\begin{abstract}
Evaluating software engineering capabilities has become a core component of modern large language models (LLMs); however, the key bottleneck hindering further scaling lies not in the scarcity of high-quality solutions, but in the lack of high-quality test suites. Test suites are indispensable both for synthesizing program repair trajectories and for providing precise feedback signals in reinforcement learning. Unfortunately, due to the high cost and difficulty of annotation, high-quality test suites have long been hard to obtain, while those automatically generated by LLMs tend to be superficial and lack sufficient discriminative power. As a first step toward constructing high-quality test suites, we introduce SWE-Mutation\blfootnote{Our code and data are available at \url{https://github.com/Sunny4Coding/SWE-Mutation}.}, a benchmark for evaluating LLM-generated test suites. The benchmark characterizes test suites by introducing systematically mutated solutions that attempt to ``fool'' the test suites and pass validation. We further propose an agentic, language-agnostic framework for automatically generating complex mutants. Our benchmark consists of 2,636 mutated variants derived from 800 original instances and includes a multilingual subset spanning nine programming languages. Experiments on seven LLMs reveal that even DeepSeek-V3.1 achieves only 10.20\% verification and 36.15\% detection rates, highlighting the inadequacy of current LLMs. Additionally, our agentic mutation strategy enhances realism, reducing average detection rates from 71.04\% to 39.81\% compared to conventional methods. These findings expose persistent deficiencies in the ability of current LLMs to generate reliable and discriminative test suites.
\end{abstract}

\section{Introduction}
\begin{figure}[t]
  \centering
  \includegraphics[width=\columnwidth]{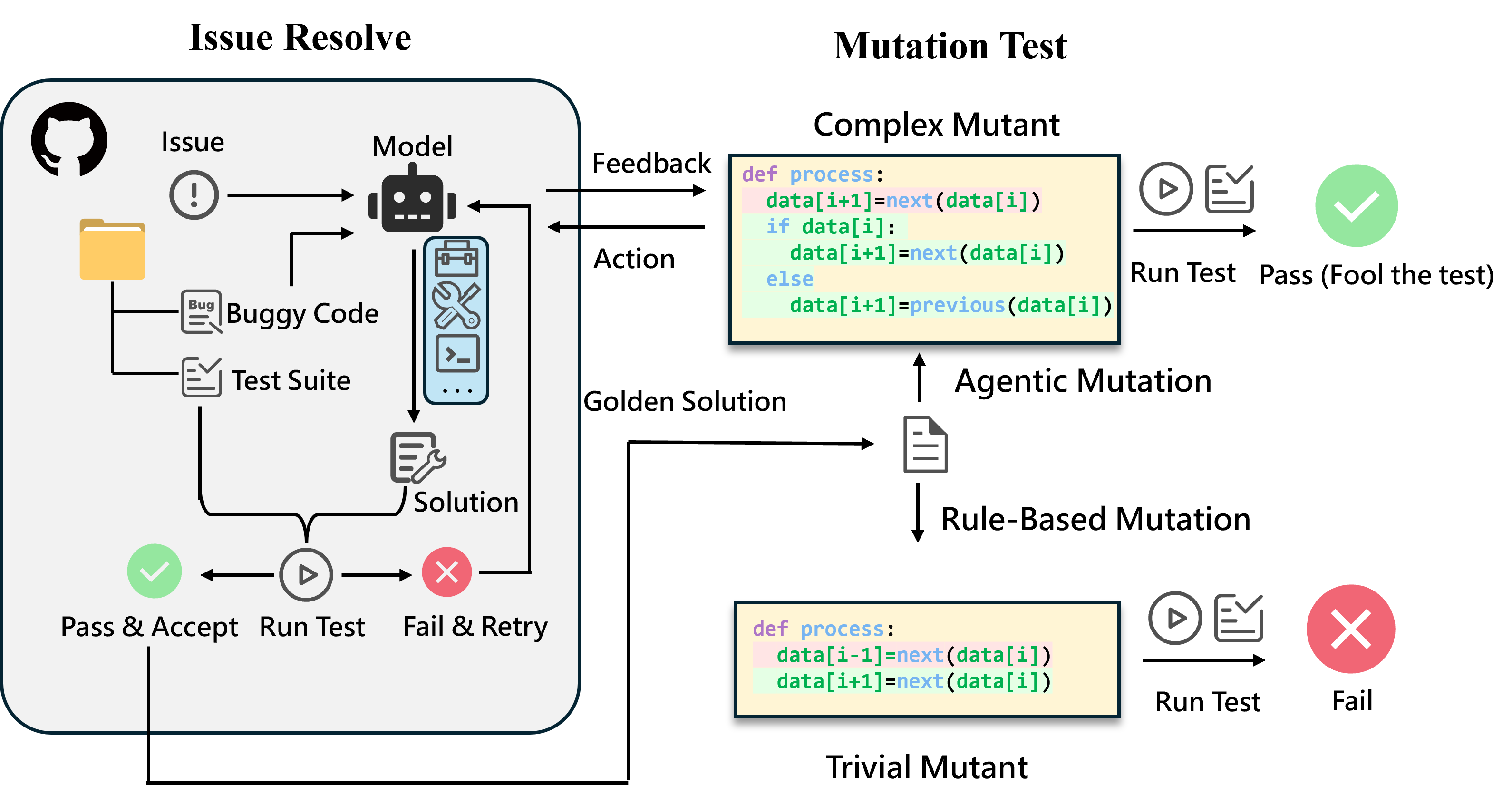}
  \caption{
    The pivotal role of test suites and high-quality mutants. 
The test suite serves as the verification standard for issue resolution. While trivial mutants are easily detected, complex mutants can successfully ``fool'' the test. This highlights that high-quality mutants are indispensable for evaluating whether a test suite is robust enough to prevent incorrect code acceptance.
  }
  \label{fig:intro}
  \vspace{-12pt}
\end{figure}

Large Language Models (LLMs) have achieved significant progress on automated software engineering (SE) tasks~\citep{jimenez2024swebench,hou2024large}.
A common paradigm for evaluating the software engineering capabilities of LLMs is to provide a software issue together with its corresponding test suite, and consider the issue solved if the model-generated solution passes all tests as illustrated in Figure~\ref{fig:intro}~\citep{chen2021evaluating,austin2021program}.
Improving the software engineering performance of LLMs---whether by synthesizing program repair trajectories for post-training or by collecting reward signals from the environment during reinforcement learning to estimate advantages---critically depends on the availability of test suites~\citep{le2022coderl,zhang2023selfedit,zhao-etal-2024-repair}.
In this sense, the ability to automatically construct high-quality, discriminative test suites would bring us substantially closer to a systematic solution to software engineering problems~\citep{chen2023codet,liu2023is}.

However, generating reliable test suites is notably challenging for LLMs. Unlike code generation, which is a constructive task aiming to produce one correct implementation, test case generation is inherently adversarial: it seeks to expose failures in an existing program by identifying rare, error-triggering inputs~\citep{yuan2023no}.
This task suffers from severe information asymmetry, extremely sparse reward signals, and a highly irregular search space, where the vast majority of inputs are uninformative.
From both theoretical and practical perspectives, test case generation is closely related to long-standing hard problems in program analysis and verification~\citep{cadar2013symbolic,king1976symbolic}, and is arguably more challenging than code generation in many realistic settings.
Therefore, as shown in Figure~\ref{fig:intro}, the synthesized test suites tend to be trivial.

As a first step toward constructing high-quality test suites, establishing effective evaluation criteria remains insufficiently explored.
Existing studies have introduced several benchmarks to assess test suite generation capabilities in software engineering tasks~\citep{liu2023is}.
These benchmarks typically characterize test suite discriminability by introducing systematically mutated, buggy solutions that attempt to ``fool'' the test suites and pass validation.
Despite substantial progress, these benchmarks still suffer from notable limitations.
In particular, the reliance on relatively homogeneous methodologies, combined with the inherent limitations of LLMs, often leads to the generation of trivial test suites that fail to meaningfully probe model capabilities.
For instance, standard approaches often employ simple rule-based operators or few-shot prompting to generate mutants~\citep{chen2023codet}.
However, previous research has questioned whether such artificial mutants are truly representative of real-world faults~\citep{just2014are,jimenez2024swebench}, as they are often easily killed by model-generated test suites as illustrated in Figure~\ref{fig:intro}.
Thus, existing benchmarks risk overestimating the quality of test suites.
Furthermore, despite the widespread adoption of agentic frameworks---which enable models to deeply understand repositories through environmental interaction~\citep{yao2022react,yang2024sweagent,10.1145/3701551.3703537}---methods for leveraging these frameworks to generate software mutants remain underdeveloped.
Second, diverse tasks require LLMs to handle multilingual repositories~\citep{cassano2023multiple,zheng2023codegeex}.
However, most existing benchmarks remain monolingual.
These constraints compromise real-world robustness and usability, posing risks to the integrity of reliable software engineering.

Recent studies have indicated that LLMs can leverage code semantics to inject subtle, realistic defects that mimic human errors~\citep{yang2025testcaseeval,11082014}. Building on this idea, we introduce \textbf{SWE-Mutation}. Our benchmark includes two tasks: test generation and test repair. Concretely, we adopt an agentic framework to generate complex mutants to reveal flaws in synthetic test suites. Each mutant represents an erroneous mutation of the golden solution in the repository and resembles realistic errors. With these mutants, our benchmark provides faithful evaluation of model abilities. By applying our framework to SWE-bench Verified~\citep{openai2024swebench},
we generated 1,664 mutants across 500 instances from 11 popular GitHub repositories. Our benchmark incorporates basic metrics like Pass$@$1 and Verified Reproduction Rate (VRR). Moreover, we introduce the Relative Detection Rate (RDR) metric, which specifically represents the relative proportion of mutants killed by the test suites. Given our language-agnostic framework, we also provide a multilingual subset with 300 instances and 972 mutants in 9 languages based on SWE-bench-Multilingual~\citep{yang2025swesmith}. 

We evaluate seven mainstream LLMs including Claude Sonnet 4.5~\citep{anthropic2025claude} and DeepSeek V3.1~\citep{deepseek2025v31}, using two agentic frameworks: Mini-Swe-Agent~\citep{yang2024sweagent} and Claude Code~\citep{anthropic2025claude}. Results show that models still struggle to generate reliable test suites. For instance, DeepSeek-V3.1 only gets 10.20\% on VRR and 36.15\% on RDR. Furthermore, models encounter significant difficulties in non-Python tasks. Our agentic mutation ensures more realistic and discriminative evaluation metrics: compared to traditional methods, the average Relative Detection Rate drops significantly from 71.04\% to 39.81\%. Finally, we provide a qualitative analysis of failure cases.


\section{Related Work}
\subsection{Benchmarks for Testing in Software Engineering}
The evaluation of LLMs in software engineering has evolved significantly. Early research primarily focused on simple code synthesis tasks using datasets like HumanEval and MBPP, where test suites served merely as verification oracles rather than generation targets~\citep{chen2021evaluating, austin2021program}. 
More recently, specific benchmarks have been developed to rigorously assess test generation abilities. For instance, TestBench and TestGenEval were introduced to evaluate the generation of unit tests and assert statements for isolated functions~\citep{zhang2025testbench, jain2025testgeneval}. 
Concurrently, to capture the complexity of real-world development, the community has established repository-level benchmarks. SWE-bench set the standard for resolving issues in Python repositories~\citep{jimenez2024swebench}, while recent extensions like SWE-smith and Multi-SWE-bench have further expanded this domain~\citep{yang2025swesmith, zan2025multiswebench}.
Despite these advancements in both unit and repository levels, existing benchmarks generally lack a robust mechanism to evaluate the quality of model-generated tests. Most rely on pass rates against human-written golden tests or simple code coverage, which are often weak proxies for assessing whether a test suite can detect subtle faults~\citep{wang2025mutation}.

\subsection{Mutation-Based Evaluation and Generation}
Mutation testing evaluates test suite quality by injecting artificial faults (mutants)~\citep{papadakis2019mutation}. Traditional generation methods rely on rule-based operators, exemplified by tools like PIT and Major~\citep{coles2016pit, just2014major}. However, these rigid rules often produce trivial mutants that fail to mimic real-world complexity~\citep{just2014mutants}. To address this, approaches such as DeepMutation and $\mu$BERT were proposed to generate more diverse mutants via neural networks~\citep{tufano2020deepmutation, degiovanni2022mubert}. Yet, these methods frequently struggle to ensure syntactic validity or semantic meaningfulness due to a lack of execution context.
Recently, LLM-based approaches have emerged as a powerful alternative. Frameworks like BugFarm and LLMorpheus leverage LLMs to generate realistic software defects~\citep{ibrahimzada2025challenging, tip2025llmorpheus}. Despite their potential, current methods predominantly rely on static prompting strategies without active environment interaction. Consequently, they often generate non-compilable code or redundant mutants that do not align with repository logic. In contrast, SWE-Mutation introduces an agentic framework that autonomously explores the repository. By analyzing code characteristics to select specific strategies, our approach generates complex semantic mutants, effectively bridging the gap between artificial mutations and realistic bugs.

\section{SWE-Mutation}
SWE-Mutation is designed to evaluate the test generation and repair abilities of LLMs within realistic software engineering scenarios. Derived from real-world GitHub repositories, the benchmark comprises 500 Python instances and 300 instances across nine other languages. In SWE-Mutation, we generated a total of 2,636 mutants. Each instance supports both tasks and is equipped with 3--5 mutants generated via our agentic framework to facilitate mutation testing. This section outlines the agentic mutation framework, task definition and characteristics of SWE-Mutation.

\subsection{Agentic Mutation Framework}
\subsubsection{Overview}
\begin{figure*}[t]
    \centering
    \includegraphics[width=0.8\linewidth]{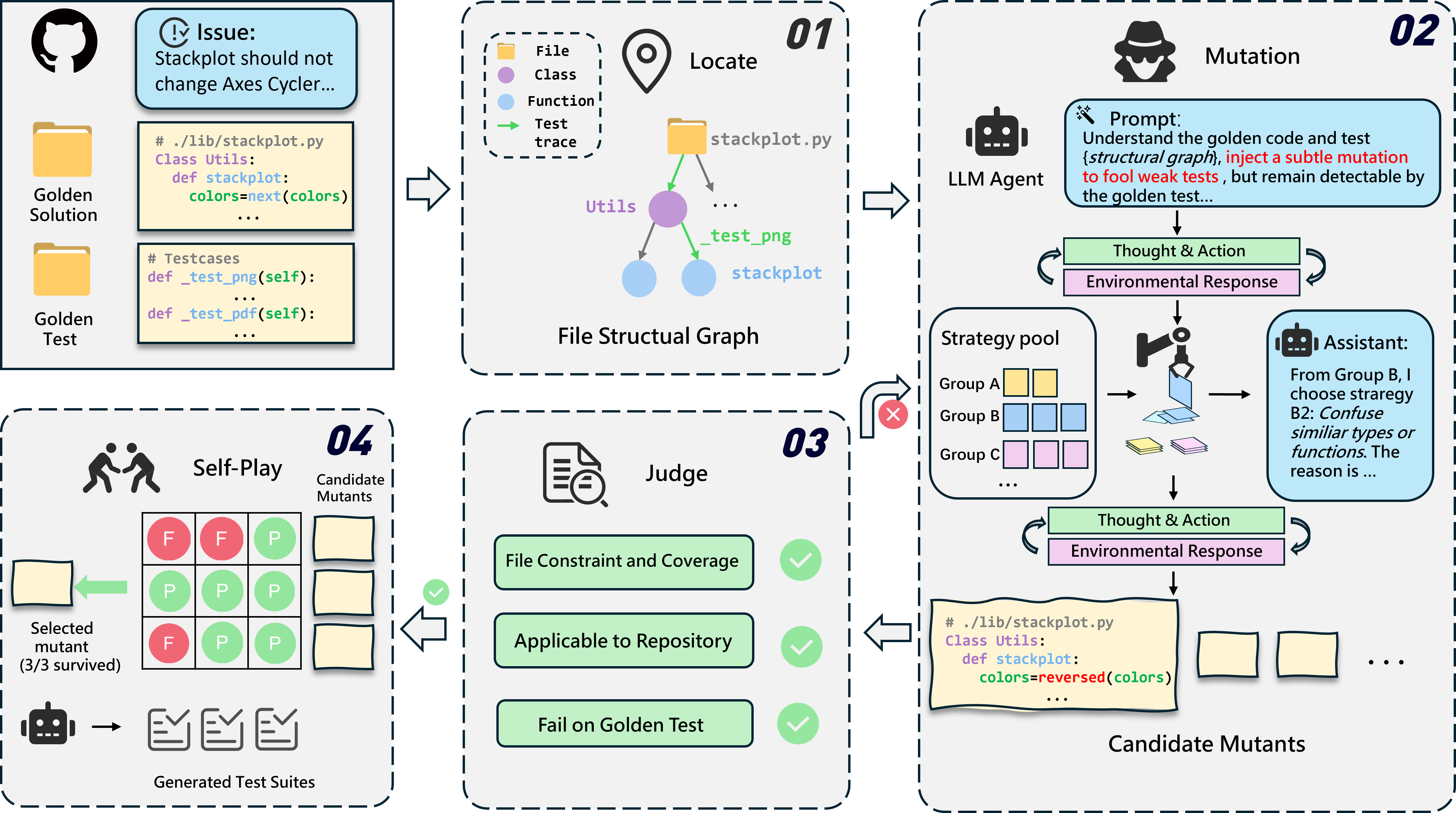}
    \caption{The overview of our framework. Starting with the golden solution and golden test suite in a repository, we employ four modules to identify scopes, generate mutants, verify validity, and perform selection.}
    \label{fig:overview}
\end{figure*}

As shown in Figure~\ref{fig:overview}, our framework to generate complex mutants includes four key modules: Locate, Mutation, Judge, and Self-Play. These modules construct a rigorous and comprehensive framework for high-quality mutant generation. Detailed statistics of the generated mutants are provided in Appendix~\ref{sec:Mutants Statistics}.

We choose the Claude Sonnet 4~\citep{anthropic2025claude} as the base model for the agent modules. To rule out a potential ``same-family'' bias from this choice, we further conduct a systematic robustness study on the full sample set ($N{=}500$) by replacing the Claude-4 generator with DeepSeek-V3.1 and Qwen3-Coder; the resulting RDR changes are within $1.5$ pp with bootstrap 95\% CIs covering zero, and the Spearman rank correlations of evaluators remain $0.96$ and $0.93$, confirming that our conclusions do not depend on the generator family (see Appendix~\ref{sec:ablation_backbone} for the full analysis). The additional details can be found in Appendix~\ref{sec:Prompts}.

\subsubsection{Locate Module}
Since our goal is to make realistic mutations rather than randomly modifying code to inject bugs, we aim to constrain code modifications within a specific scope~\citep{yu-etal-2025-testagent}. In the Locate module, we restrict the files available for mutation to those modified in the golden solution. Additionally, we use Tree-sitter to parse these files and extract execution traces from Fail-to-Pass tests in the golden test suite. This is because Fail-to-Pass (F2P) tests fail on buggy code and pass on fixed code, identifying the specific defect. We focus on F2P because it uniquely captures the bug's triggering logic and verifies the correctness of the fix. By annotating this trace onto the structural graph, we assist the model in understanding call relationships and test logic, enabling precise mutation.

\subsubsection{Mutation Module}
We aim to generate mutants that are realistic and difficult for test suites to kill. To do this, we analyzed common errors models make in SWE tasks. These errors often evade model-generated tests, leading to fix failures. We categorize these errors into five strategies. Detailed descriptions and examples for each strategy are provided in Appendix~\ref{sec:Strategies in Mutation Module}. In each run, we select one strategy group. The model analyzes the tests in the golden test suite. It modifies code within the scope defined by the Locate module. Then, the model injects bugs based on the chosen strategy. To ensure quality and interpretability, we require the model to provide reasoning for its modifications. Examples of each strategy are provided in Appendix~\ref{sec:Case Studies}.

\subsubsection{Judge Module}
We establish the Judge module to ensure the rationality of generated mutants. In this module, we validate the mutants against the golden test suite under three strict constraints to ensure they resemble realistic, common errors. First, the modifications must be restricted to the files touched by the golden solution. Second, the mutant must be successfully applied to the repository and pass syntax or compilation checks. Third, it must fail at least one F2P test. Instances that do not meet these requirements are returned to the Mutation module for revision within retry limits.

\subsubsection{Self-Play Module}
We construct the Self-Play module to ensure that the generated mutants have difficulty and discriminatory power. Within this module, we apply a selection procedure to eliminate trivial mutants. First, leveraging the Mutation module, we sample $N$ diverse candidate mutants by applying varying mutation strategies. Second, we use the model to generate 10 test suites for each instance under the original task setting with temperature. Third, we evaluate each candidate against these generated test suites. A high survival rate indicates that the mutant has successfully evaded detection by the model. We rank and select the top 50\% that successfully evaded more than 3 test suites.

\subsection{Task Formulation}
For each instance in the SWE-Mutation benchmark, we design two tasks to evaluate the abilities of LLMs: test generation and test repair.

\textbf{Test Generation.} The goal of the test generation task is to generate a complete test suite from scratch for a given problem. In this process, we only provide the model with the file path where the test suite should be generated. The model is then expected to complete the test suite by its comprehensive understanding of the code repository.

\textbf{Test Repair.} The goal of the test repair task is to fix an existing test suite that is currently incomplete or flawed (i.e., it fails to detect existing bugs in the current repository). This repair can be achieved by adding new test functions or modifying existing ones. Compared to generating tests entirely from scratch, this task more closely reflects real-world SE scenarios.  

For both tasks, we expect the generated tests to effectively detect the issues in the original buggy repository without incorrectly failing the golden solution. Furthermore, a high-quality test suite should identify as many mutants as possible. This can demonstrate the robustness of the model's generated tests. The specific prompts used for these tasks can be found in Appendix~\ref{sec:Prompts}.

\subsection{Benchmark Characteristics}
We detail the features that distinguish SWE-Mutation from existing test suite benchmarks. This is also illustrated in Table~\ref{tab:characteristics}.

\newcommand{\cmark}{\ding{51}}
\newcommand{\xmark}{\ding{55}}

\begin{table*}[t]
  \centering
  \caption{
Comparison of SWE-Mutation with state-of-the-art benchmarks. 
  SWE-Mutation distinguishes itself to integrate agentic mutation within a repository-level environment across 10 languages, addressing the limitations of static generation and monolingual focus in prior works.
  }
  \label{tab:characteristics}
  
  \resizebox{\textwidth}{!}{
    \begin{tabular}{l c c c l}
      \toprule
      \textbf{Benchmark} & \textbf{Scale} & \textbf{Langs} & \textbf{Agentic Loop} & \textbf{Mutation} \\
      \midrule
      
      TestEval~\citep{wang2025testeval} & Function & Python & \ding{55} & None \\
      LLMorpheus~\citep{tip2025llmorpheus} & Function & JavaScript & \ding{55} & Few-shot LLM  \\
      BugFarm~\citep{ibrahimzada2025challenging} & Function & Java & \ding{55} & Pipeline LLM  \\
      TestGenEval~\citep{jain2025testgeneval} & File & Python & \ding{55} & Rule-based \\
      SWT-bench~\citep{mundler2024swt} & Repository & Python & \ding{51} & None \\
      
      \midrule
      
      \textbf{SWE-Mutation (Ours)} & \textbf{Repository} & \textbf{10 languages} & \textbf{\ding{51}} & \textbf{Agentic Framework} \\
      \bottomrule
    \end{tabular}
  }
\end{table*}

\textbf{Repository-Level Environment:} Unlike benchmarks restricted to file-level snippets, ours is built on well-maintained GitHub repositories with complete execution environments. Models interact via tools and CLIs, simulating complex SE scenarios rather than isolated coding tasks.

\textbf{Support for Agentic Frameworks:} While existing benchmarks rely on prompt-only interactions, SWE-Mutation is specifically designed for agentic workflows. It supports popular frameworks like Mini-Swe-Agent and Claude Code, enabling the evaluation of agentic abilities.

\textbf{Multi-language Benchmark:} Addressing the Python-centric limitation of current studies, we introduce SWE-Mutation-Multilingual. This extension supports nine programming languages: C, C++, Java, TypeScript, JavaScript, Rust, Go, PHP and Ruby, filling the void for repository-level evaluation in diverse linguistic settings.

\textbf{Agentic Mutation:} Conventional methods often rely on trivial rule-based mutations that lack semantic context and are easily detected by advanced models. In contrast, our benchmark employs an agentic framework to synthesize complex mutants. This approach provides a significantly more robust metric for evaluating model abilities.

\section{Experiments}

\subsection{Models} 
For models, we evaluate seven mainstream open-source and closed-source models, including Claude Sonnet 4.5 and Claude Sonnet 3.7~\citep{anthropic2025claude},
Qwen3-Coder-480B-A35B-Instruct~\citep{qwen3technicalreport}, DeepSeek-V3.1~\citep{deepseek2025v31}, Kimi K2-0905~\citep{kimiteam2025kimik2openagentic}, GPT-oss-120B~\citep{openai2025gptoss120bgptoss20bmodel},
and GLM-4.6~\citep{zhipuai2025glm46}. For brevity, some model names will be abbreviated throughout this paper.

\subsection{Agentic Systems} 
Due to the complexity of test-suite tasks, directly prompting LLMs to produce the required changes is not effective. Therefore, we use two advanced agentic tools to address these tasks: the open-source Mini-Swe-Agent~\citep{yang2024sweagent} and the proprietary Claude Code~\citep{anthropic2025claude}. They represent two different approaches to automating software engineering tasks.

Mini-Swe-Agent enables LLMs to solve problems by interacting with codebases solely through bash commands. We select it for its lightweight design, consisting of only about 100 lines of Python code, which makes it easy to deploy. It outputs a complete shell command at each step without relying on separate ``tool call'' protocols.

Claude Code is a command-line interface (CLI) tool designed specifically for Claude models. It allows the model to perform software tasks directly in the terminal. Unlike the general approach of Mini-Swe-Agent, Claude Code uses an optimized workflow for Claude to read, edit, and run code, representing a highly integrated agentic solution.

\subsection{Evaluation Metrics} 
\label{sec:metrics}

To evaluate the generated test suites effectively, we use three metrics: Pass$@$1, Verified Reproduction Rate (VRR) and Relative Detection Rate (RDR). The definition is detailed below:

Pass$@$1:
We employ Pass$@$1 to evaluate the model's success rate in generating valid test cases within a single attempt. A generated test suite is considered successful if it is a correctly formatted git diff and can be successfully applied to the repository. Also, it should execute without triggering compilation errors.

Verified Reproduction Rate (VRR):
VRR measures how often the model successfully reproduces the specific issue without breaking correct functionality. A generated test suite is considered a success only if it satisfies two conditions. 1): Reproduction: It fails on the original bug. 2): Validity: It passes on the fixed golden code. We define VRR as the proportion of instances in the dataset where the generated test suite satisfies both validity and reproduction rules. This metric ensures that we only credit tests that demonstrate both correctness and effectiveness.

Relative Detection Rate (RDR):
RDR evaluates the effectiveness of test suites in detecting mutants. Let $\mathcal{M}^{(i)}$ denote the total set of mutants generated for instance $i$. We define two subsets:

 $M_{base}^{(i)}$: The set of mutants killed by test suite in the original repository.
 
$M_{gen}^{(i)}$: The set of mutants killed by the model-generated test suite.

To focus on evaluating the model's performance on the specific subset of mutants that the original test suite failed to kill, we formulate RDR using the set difference operation:

\begin{equation} \label{eq:rdr}
    \text{RDR} = \frac{\sum_{i=1}^{N} \left| M_{gen}^{(i)} \setminus M_{base}^{(i)} \right|}{\sum_{i=1}^{N} \left| \mathcal{M}^{(i)} \setminus M_{base}^{(i)} \right|}
\end{equation}

Equation~\ref{eq:rdr} explicitly computes a micro-average by summing the absolute counts of total surviving mutants (denominator) and those successfully killed by the model (numerator) across all instances. This ensures each mutant contributes equally, preventing instances with very few mutants from disproportionately skewing the overall metric. To quantify the reliability of RDR, we further report instance-level bootstrap 95\% confidence intervals ($10{,}000$ resamples) and non-parametric Wilcoxon signed-rank tests between the top model and each competitor; the detailed statistics are provided in Appendix~\ref{sec:rdr_stats}.

Test repair ($M_{base} \neq \varnothing$): 
    The metric evaluates the incremental value. It strictly focuses on the new mutants killed by the model that were missed by the original developers.
    
Test generation ($M_{base} = \varnothing$): 
    In the absence of an original test suite, the set difference simplifies (i.e., $\mathcal{M} \setminus \varnothing = \mathcal{M}$). In this case, the metric reduces to the absolute Mutation Score ($\frac{|M_{gen}|}{|\mathcal{M}|}$), measuring the model's overall ability.

\subsection{Main Results}
\label{sec:results}
\begin{table*}[ht]
  \centering
  \small
  \renewcommand{\arraystretch}{1.2}
  \setlength{\tabcolsep}{4pt}
  \newcolumntype{Y}{>{\centering\arraybackslash}X}

  \caption{\label{tab:test_repair}
    Performance comparison on test repair task across different LLMs on SWE-Mutation.
  }
  \begin{tabularx}{\textwidth}{l *{6}{Y}}
    \toprule
    \multirow{2}{*}{Model} & \multicolumn{3}{c}{Mini-Swe-Agent} & \multicolumn{3}{c}{Claude Code} \\
    \cmidrule(lr){2-4} \cmidrule(lr){5-7}
    & Pass$@$1(\%) & VRR(\%) & RDR(\%) & Pass$@$1(\%) & VRR(\%) & RDR(\%) \\
    \midrule
    Claude-sonnet-4.5 & 97.20 & 42.60 & 79.30 & 99.80 & 59.20 & 81.15 \\
   Claude-sonnet-3.7 & 94.40 & 29.80 & 62.58 & 97.60 & 52.80 & 66.59 \\
    DeepSeek-V3.1 & 96.60 & 33.00 & 66.41 & 96.80 & 58.20 & 68.36\\
    Qwen3-Coder & 87.60 & 38.00 & 68.99 & 96.40 & 50.00 & 70.21 \\
   Kimi-K2 & 83.80 & 40.60 & 74.58 & 86.00 & 54.40 & 74.19 \\
    GLM-4.6 & 83.40 & 29.40 & 71.26 & 95.60 & 49.80 & 73.54 \\
   GPT-oss-120B & 74.80 & 24.80 & 36.31 & 86.80 & 36.40 & 39.28 \\
    \bottomrule
  \end{tabularx}

  \vspace{0.5cm}

  \caption{\label{tab:test_generation}
    Performance comparison on test generation task across different LLMs on SWE-Mutation.
  }
  \begin{tabularx}{\textwidth}{l *{6}{Y}}
    \toprule
    \multirow{2}{*}{Model} & \multicolumn{3}{c}{Mini-Swe-Agent} & \multicolumn{3}{c}{Claude Code} \\
    \cmidrule(lr){2-4} \cmidrule(lr){5-7}
    & Pass$@$1(\%) & VRR(\%) & RDR(\%) & Pass$@$1(\%) & VRR(\%) & RDR(\%) \\
    \midrule
    Claude-sonnet-4.5 & 96.20 & 29.80 & 63.70 & 98.00 & 40.40 & 71.71 \\
    Claude-sonnet-3.7 & 88.40 & 20.60 & 37.47 & 95.40 & 28.80 & 38.60 \\
    DeepSeek-V3.1 & 88.20 & 10.20 & 36.15 & 94.00 & 20.40 & 39.09 \\
    Qwen3-Coder-480B & 86.20 & 12.40 & 33.33 & 95.20 & 26.80 & 33.21 \\
    Kimi-K2 & 79.40 & 14.60 & 42.59 & 83.60 & 19.20 & 45.12 \\
    GLM-4.6 & 74.60 & 15.20 & 39.79 & 86.20 & 25.40 & 42.11 \\
    GPT-oss-120B & 59.80 & 8.00 & 25.61 & 65.60 & 19.20 & 28.73 \\
    \bottomrule
  \end{tabularx}
\end{table*}
We evaluate seven state-of-the-art LLMs on SWE-Mutation using two agentic frameworks. The results on test generation and test repair tasks are presented in Table~\ref{tab:test_repair} and Table~\ref{tab:test_generation}. Our analysis reveals that even the most advanced models and frameworks exhibit limited performance. Specifically, Claude-sonnet-4.5 consistently achieves the best performance across all metrics and tasks, significantly outperforming other models. For instance, in the test repair task under the Claude Code framework, it attains the highest VRR (59.20\%) and RDR (81.15\%). Notably, while most models achieve high Pass$@$1 scores, their VRR scores remain significantly lower. This discrepancy indicates that while models can easily generate syntactically correct code, constructing logically correct tests that reproduce bugs remains a challenge.

For task difficulty, test generation proves to be significantly more challenging than test repair. Comparing the two tables, we observe a universal decline in metrics for the generation task. For example, even for the top-performing Claude-sonnet-4.5, the VRR drops from 42.60\% (Repair) to 29.80\% (Generation) under the Mini-Swe-Agent framework. This underscores the higher complexity involved in synthesizing complete test suites from scratch compared to fixing existing ones.

\begin{figure}[t]
  \centering
  \includegraphics[width=\columnwidth]{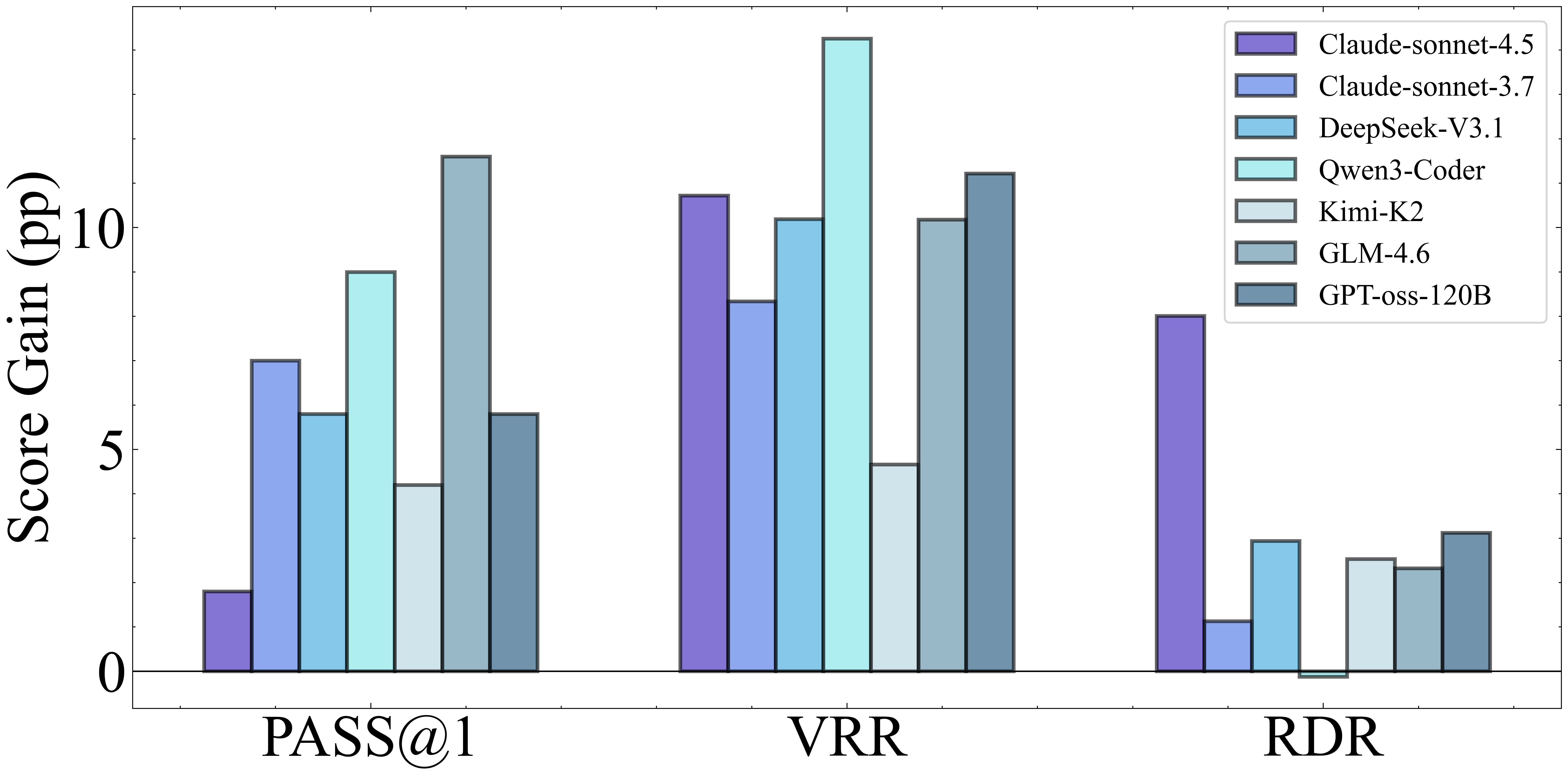}
  \caption{\label{fig:score_gain}Performance gains achieved by switching from Mini-Swe-Agent to Claude Code.
  }
  \vspace{-20pt}
\end{figure}

Regarding the framework impact, Claude Code demonstrates superior effectiveness compared to Mini-Swe-Agent. Detailed log analysis reveals that Mini-Swe-Agent relies solely on simple bash commands. When handling the extensive code writing for test generation, the use of ``sed'' commands leads to indentation errors and other syntax issues. In contrast, Claude Code addresses this with specialized ``Edit'' tools, which facilitate the generation and modification of large-scale files. Interestingly, as illustrated in Figure~\ref{fig:score_gain}, while models running on Claude Code generally achieve higher Pass$@$1 and VRR scores, there is no significant upward trend in RDR scores. This suggests that switching frameworks primarily enhances basic ability, such as formatting correctness and reproducing the bug. However, detecting complex semantic mutants requires a profound understanding of repository context, so the model's inherent ability remains the dominant factor~\citep{zhao2025unveiling}.

\subsection{Results on SWE-Mutation-Multilingual}
\begin{figure}[t]
  \centering
  \includegraphics[width=\columnwidth]{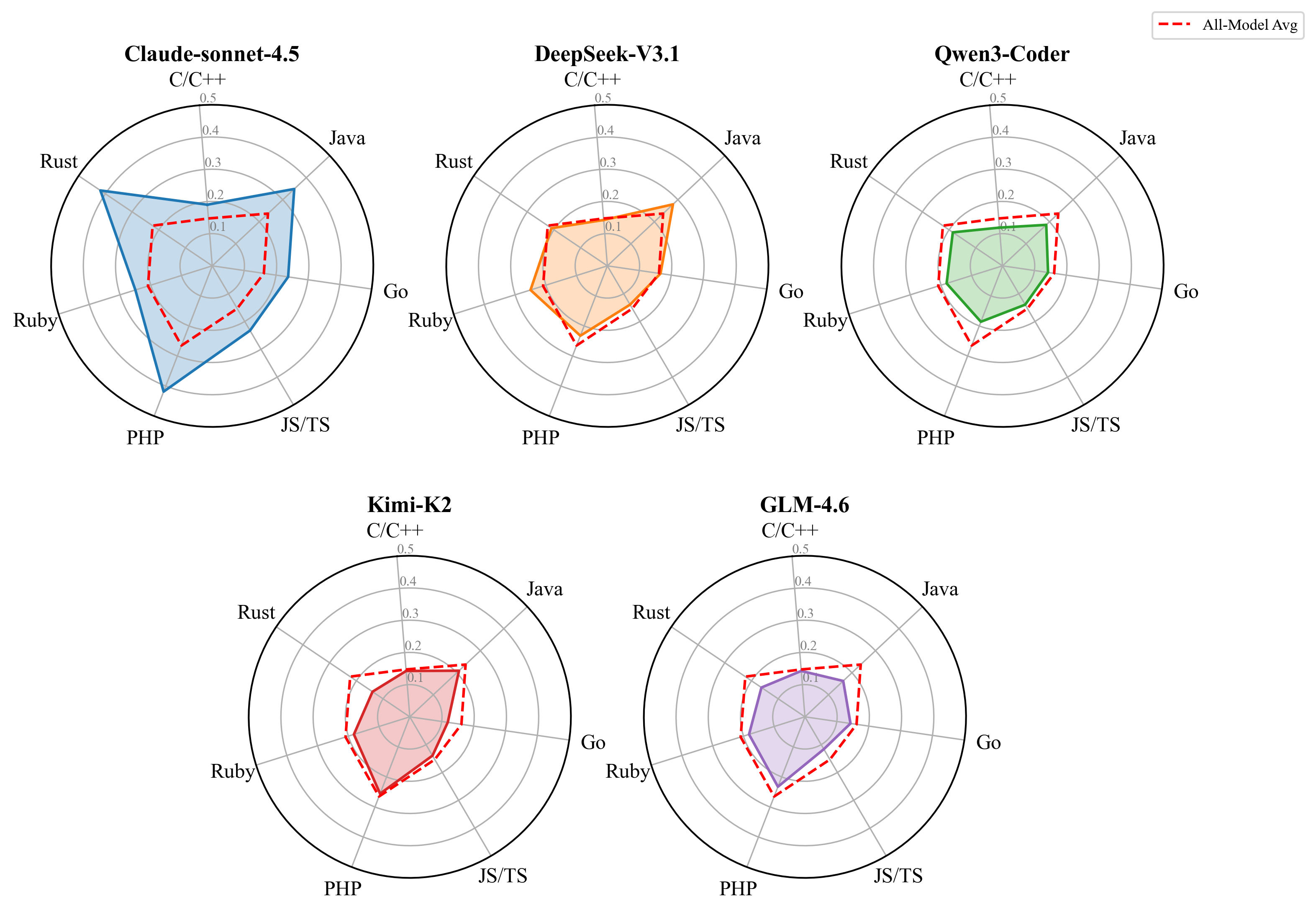}
  \caption{
    RDR performance across 9 programming languages for different LLMs.
  }
  \label{fig:multilang_radar}
\end{figure}

\begin{figure*}[h]
  \centering
  \includegraphics[width=\textwidth]{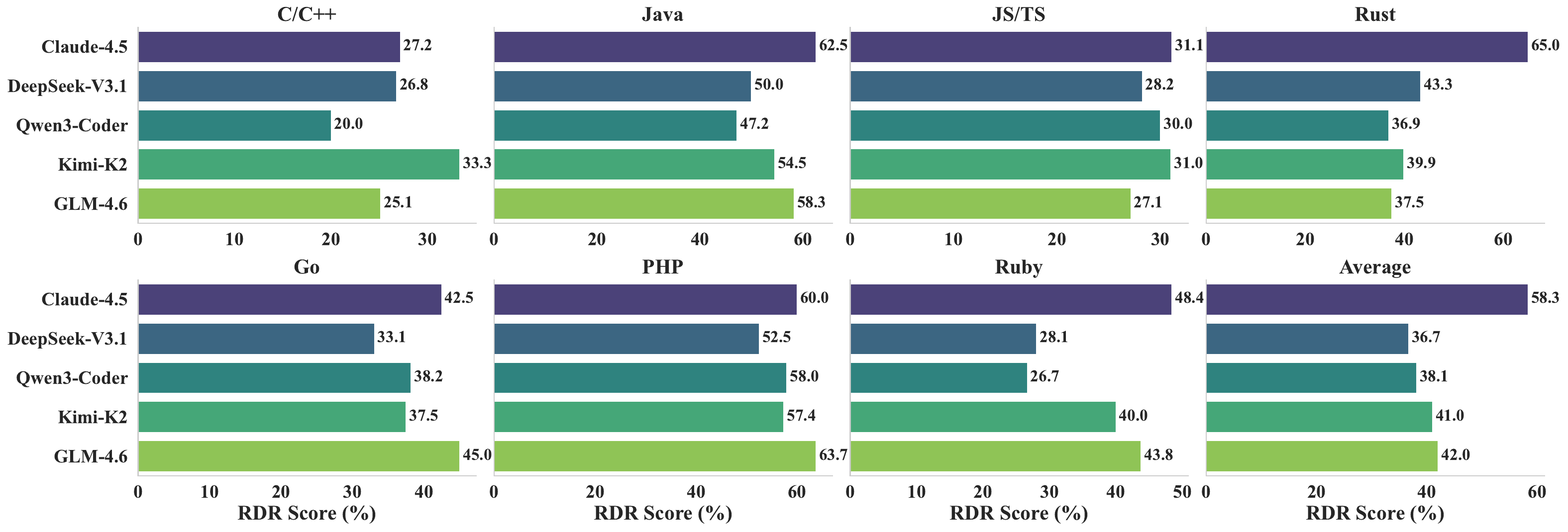}
  \caption{
    VRR performance across 9 programming languages for different LLMs.}
  \label{fig:multilang_facet}
  \vspace{-8pt}
\end{figure*}

\begin{table}[ht]
  \centering
  \small
  \renewcommand{\arraystretch}{1.2}
  \setlength{\tabcolsep}{3.5pt}
  \caption{\label{tab:multilang_results}
    Performance of LLMs on test repair task in SWE-Mutation-Multilingual (Mini-Swe-Agent). Results are averaged across 9 programming language instances.
  }
  \begin{tabular}{l ccc}
    \toprule
    Model & Pass$@$1 (\%) & VRR (\%) & RDR (\%) \\
    \midrule
    Claude-sonnet-4.5 & 91.33 & 33.33 & 58.33 \\
    DeepSeek-V3.1     & 86.00 & 20.33 & 36.67 \\
    Qwen3-Coder       & 83.67 & 16.33 & 38.13 \\
    Kimi-K2           & 80.67 & 17.00 & 41.00 \\
    GLM-4.6           & 81.67 & 15.33 & 42.05 \\
    \bottomrule
  \end{tabular}
\vspace{-4pt}
\end{table}
We evaluate the test repair task across 9 programming languages with Mini-Swe-Agent as the framework. Table~\ref{tab:multilang_results} details the performance of five models on the test repair task. First, performance drops significantly compared to Python. Claude-sonnet-4.5 achieves a VRR of 42.60\% and an RDR of 79.30\% on Python tasks. However, in the multi-language setting, the scores drop to 33.33\% and 58.33\%, respectively. Other models show even steeper declines. This confirms that non-Python software engineering tasks are much more challenging for current LLMs. Additionally, performance varies significantly across languages. We visualize the detailed VRR and RDR distributions in Figure~\ref{fig:multilang_radar} and~\ref{fig:multilang_facet}. As shown, Claude-sonnet-4.5 demonstrates robust generalization across diverse languages. In contrast, other models show contracted and irregular shapes. Specifically, performance on Java, PHP, and Rust significantly outperforms that on C/C++ and JS/TS. Our analysis identifies the root cause. Models encounter major hurdles when synthesizing tests involving memory management (typical in C/C++) and event-driven mechanisms (typical in JS/TS). In Appendix~\ref{sec:Case Studies}, we provide an analysis of representative failure cases.

\subsection{Comparison between Mutation Strategies}
We compare three mutation strategies on the test generation task: rule-based mutation~\citep{jain2025testgeneval}, few-shot LLM~\citep{tip2025llmorpheus} and our agentic semantic mutation. Specific prompt settings can be found in Appendix~\ref{sec:Mutation Methods Settings}. In this context, since there is no baseline test suite ($M_{base} = \varnothing$), the RDR metric degenerates to the absolute percentage of mutants killed. As shown in Table~\ref{tab:mutation_comparison}, models achieve the highest scores on rule-based mutants. This indicates that random syntax errors are trivial to detect. Scores drop partially on mutants generated by few-shot LLM. However, models obtain the lowest scores on our semantic-level mutation. For instance, the score of Claude-sonnet-4.5 drops from 85.40\% to 63.70\%. This proves that our semantic mutants are much harder to kill. They effectively expose the limitations of generated tests. 
\begin{table}[ht]
  \centering
  \small
  \renewcommand{\arraystretch}{1.2}
  \setlength{\tabcolsep}{3.5pt}
  \caption{\label{tab:mutation_comparison}
    Comparison of RDR scores across different mutation strategies on the test generation task.
  }
  \begin{tabular}{l ccc}
    \toprule
    \multirow{2}{*}{Model} & \multicolumn{3}{c}{RDR on Mutation Strategy (\%)} \\
    \cmidrule(lr){2-4}
    & Rule-Based & Few-shot & \textbf{Agentic (Ours)} \\
    \midrule
    Claude-sonnet-4.5 & 75.43 & 69.52 & \textbf{63.70} $\downarrow$ \\
    Claude-sonnet-3.7 & 73.25 & 55.25 & \textbf{37.47} $\downarrow$ \\
    DeepSeek-V3.1     & 72.92 & 52.86 & \textbf{36.15} $\downarrow$ \\
    Qwen3-Coder       & 72.16 & 50.18 & \textbf{33.33} $\downarrow$ \\
    Kimi-K2           & 74.12 & 62.43 & \textbf{42.59} $\downarrow$ \\
    GLM-4.6           & 73.88 & 59.55 & \textbf{39.79} $\downarrow$ \\
    GPT-oss-120B      & 55.55 & 35.27 & \textbf{25.61} $\downarrow$ \\
    \bottomrule
  \end{tabular}
\end{table}
We further analyze the score distribution across strategies. For rule-based method, model scores remain tightly clustered within a narrow range. This convergence suggests that syntactic mutants fail to distinguish model capabilities. In contrast, our method reveals distinct performance gaps between models. This variance confirms that our strategy offers superior discrimination. Additionally, evaluations using standard SE metrics further validate the quality of our mutants. Detailed results are available in Appendix~\ref{sec:Mutants Statistics}.

\subsection{Failure Analysis}
\label{sec:failure_analysis}
To understand why models fail on SWE-Mutation tasks, we manually checked the failed instances. We summarize the main reasons below.

\textbf{Lack of Global Understanding:} Models tend to focus on local file content and overlook global structures, such as class hierarchies or project utilities. This leads to calls to undefined methods or the misuse of internal APIs. Furthermore, models struggle with environment dependencies and frequently confuse relative and absolute import paths or hallucinate non-existent libraries. These errors directly cause execution failures.

\textbf{Instability in Cross-file Interaction and Long Contexts:} Models struggle to track data flow across multiple files and often fail to instantiate objects defined in separate directories. Moreover, generating extensive test files is error-prone. Models lose coherence during long code generation. We observed frequent indentation errors and truncated code. This issue is particularly severe when using simple command-line tools. We provide a detailed analysis of this in Appendix~\ref{sec:Case Studies}.

\textbf{Cross-Lingual Failure Patterns:} To further characterize how failure modes vary across programming languages, we manually inspect the failed instances on SWE-Mutation-Multilingual and summarize the most frequently observed error patterns for each of the nine languages. As shown in Table~\ref{tab:cross_lingual_failures} (Appendix~\ref{sec:cross_lingual_failures}), while the dominant failure types naturally differ across languages, \emph{issue misunderstanding} and \emph{inconsistent environment/package versions} remain the two most pervasive causes of failure globally. Note that because tests generated under instances sharing identical third-party libraries (e.g., \texttt{astropy} in Python and \texttt{lombok} in Java) often exhibit common, library-specific defects, and the distribution of these libraries across instances is highly uneven, we refrain from reporting strict quantitative statistics so as to avoid this confounding factor. This finding suggests that, beyond language-specific tooling issues, models must pay closer attention to comprehensive repository-level semantic understanding and precise environment version states when generating test cases.
\section{Conclusion}
We introduced SWE-Mutation, a benchmark leveraging agentic frameworks to generate complex semantic mutants for evaluating test suite robustness. Unlike trivial syntactic baselines, our approach creates realistic errors that reveal severe deficiencies in current LLMs. For instance, DeepSeek-V3.1 achieves only 10.20\% Verified Reproduction Rate (VRR) and 36.15\% Relative Detection Rate (RDR), with performance further degrading in multilingual settings. By demonstrating superior discriminability, SWE-Mutation serves as a rigorous testbed for autonomous software engineering, guiding future work toward enhanced reasoning and diverse mutation strategies.
\section*{Limitations}
In this section, we discuss potential limitations inherent to the SWE-Mutation benchmark.

\textbf{Dependence on Data Quality: }The reliability of SWE-Mutation fundamentally rests on the quality of the real-world repositories. Our evaluation framework assumes that the provided ``golden solutions'' and ``golden test suites'' serve as the absolute ground truth. However, even in the highly maintained repositories included in SWE-bench, the code may not be exhaustive or entirely error-free. This phenomenon, widely known in software testing as the \textit{oracle problem}~\citep{ebetal:oracle}, implies that potential imperfections in the ground truth could introduce biases into the evaluation.

\textbf{Fairness: }We employ Claude-4 for mutant generation within SWE-Mutation. We acknowledge that this could introduce a ``same-family'' bias; for example, Claude-3.7 might exhibit performance anomalies on artifacts generated by its predecessor due to distributional alignment. To address this, we conduct a systematic control experiment on the full sample set ($N=500$) by replacing the default Claude-4 generator with DeepSeek-V3.1 and Qwen3-Coder, and re-evaluate seven representative Claude and non-Claude models, reporting bootstrap 95\% CIs for RDR and its change $\Delta$, as well as Spearman rank correlations across backbones. We find that all $\Delta$ CIs cover zero and rank correlations remain high ($\rho=0.96$ and $0.93$), indicating that the relative rankings are robust and no additional Claude-family advantage is statistically detectable. These findings are discussed in detail in Appendix~\ref{sec:ablation_backbone}.

\section*{Acknowledgments}
This work was supported by the National Natural Science Foundation of China (Grants No.62477044 and 62406303), the Key Technologies R\&D Program of Anhui Province (No.202423k09020039), the Young Elite Scientists Sponsorship Program by CAST (No.2024QNRC001), the Fundamental Research Funds for the Central Universities (No.WK2150110038), and the Open Project of the Key Laboratory of Data Intelligence and Advanced Computing in Provincial Universities, Soochow University (KJS2410).


\bibliography{custom}

\appendix

\section{Strategies in Mutation Module}
\label{sec:Strategies in Mutation Module}
In this section, we detail the mutation strategies employed in the Mutation module. We categorize these strategies into five groups, each comprising specific sub-types. During generation, the model autonomously selects a sub-type from a group to synthesize mutants iteratively. Derived from empirical observations of real-world human errors and common model failures in SWE tasks, these strategies are designed with multilingual extensibility to support diverse programming environments.
\setminted{
    fontsize=\scriptsize,    
    breaklines=true,         
    breakanywhere=true,      
    frame=lines,             
    framesep=2mm             
}

\subsection*{a: Violation of API Specifications \& Contracts}
This strategy alters expected API behaviors. It includes modifying parameter defaults, swapping orders, or changing exception types. The code remains syntactically correct. However, it breaks semantic contracts. Surface-level tests easily miss these violations.

\subsubsection*{a1: Alter Parameter Default or Semantics}
Modify a function's default parameter to a subtle but valid edge case, or alter the internal logic handling that parameter.

\begin{listing}[H]
\input{16055F7D710165A23947CA79640B9554.highlight.minted}
\caption{a1: Altering default timeout parameters}
\label{code:a1}
\end{listing}

\subsubsection*{a2: Break API Signature or Convention}
Swap the order of function parameters, causing failures for positional-argument-based tests while likely passing keyword-argument-based ones.

\begin{listing}[H]
\input{5896A16936F248A5D5A342461A447470.highlight.minted}
\caption{a2: Swapping homogeneous parameters}
\label{code:a2}
\end{listing}

\subsubsection*{a3: Substitute Exception or Warning Type}
Replace a specific exception type with another logically related one, to fool generic \texttt{except Exception} blocks while failing precise \texttt{except SpecificError} checks.

\begin{listing}[H]
\input{274CDD982FC0481296BC2563328BE3A6.highlight.minted}
\caption{a3: Relaxing exception contracts (Silent Failure)}
\label{code:a3}
\end{listing}

\subsubsection*{a4: Violation of Read-Only Contracts}
Introduce side-effects into methods that are semantically designed for query or validation purposes (e.g., getters or validators). This breaks the ``Command-Query Separation'' principle, causing data corruption when innocent read operations are performed.

\begin{listing}[H]
\input{B8BDA6155F6CAC7001E1AEB34084C991.highlight.minted}
\caption{a4: Implicit side-effects in read-only methods}
\label{code:a4}
\end{listing}

\subsection*{b: Manipulation of Boundaries \& Conditional Logic}
This strategy introduces subtle tweaks to logic. Examples include off-by-one errors, removing null checks, or inverting booleans. Typical inputs pass successfully. Failures only occur in edge cases. Consequently, conventional tests struggle to kill them.

\subsubsection*{b1: Introduce Off-by-One Boundary Error}
Change a comparison operator like \texttt{>=} to \texttt{>}, which passes tests using typical values but fails on the exact boundary value.

\begin{listing}[H]
\input{D754602B538E9166BB5D637B40A6D424.highlight.minted}
\caption{b1: Off-by-one error in batch processing}
\label{code:b1}
\end{listing}

\subsubsection*{b2: Remove Null/Empty Case Handling}
Delete or comment out pre-condition checks for \texttt{None} or empty collections, causing downstream errors that simple tests might not trigger.

\begin{listing}[H]
\input{F28ADB1F30DECC853184A6D61DA1838B.highlight.minted}
\caption{b2: Removing safety guards for null inputs}
\label{code:B2}
\end{listing}

\subsubsection*{b3: Invert Boolean Logic or Comparison}
Flip boolean operators like \texttt{and} to \texttt{or}, which may not be caught by tests that only check all-true or all-false input combinations.

\begin{listing}[H]
\input{01CCAC54C896AFC0B78A1FDAFCC7F2FC.highlight.minted}
\caption{b3: Weakening boolean logic in security checks}
\label{code:b3}
\end{listing}

\subsection*{c: Alteration of Type \& Data Shape}
This strategy changes type handling or precision requirements. It breaks implicit type coercion or reduces numerical accuracy. These modifications primarily affect flexibility. Single-type checks or low-precision tests often fail to detect these shifts.

\subsubsection*{c1: Break Implicit Type Coercion}
Remove code that normalizes or converts inputs into a standard type, causing failures for inputs that rely on this implicit flexibility.

\begin{listing}[H]
\input{8E4A9C0930BE5754A5792E9E51613836.highlight.minted}
\caption{c1: Removing type flexibility}
\label{code:c1}
\end{listing}

\subsubsection*{c2: Reduce Numerical Precision}
Replace a high-precision numerical operation, like \texttt{math.isclose}, with standard floating-point logic (\texttt{==}) that fails due to precision errors.

\begin{listing}[H]
\input{B4C6894873F3A27FA8C557FB94697B93.highlight.minted}
\caption{c2: Downgrading numerical precision}
\label{code:c2}
\end{listing}

\subsubsection*{c3: Confuse Text vs. Bytes Encoding}
Remove an explicit \texttt{encoding} argument from a file or network I/O operation, making it rely on an unstable system default.

\begin{listing}[H]
\input{0F5210B35A0F6AA37927D5B792970533.highlight.minted}
\caption{c3: Ignoring explicit encoding requirements}
\label{code:c3}
\end{listing}

\subsection*{d: Violation of Stateful Logic \& Sequences}
This strategy disrupts object states or call sequences. It causes issues like incomplete initialization or broken idempotency. Single calls function correctly. However, multi-step or state-dependent scenarios expose the flaws.

\subsubsection*{d1: Break State Initialization or Reset}
Modify \texttt{\_\_init\_\_} or a \texttt{reset()} method to incompletely initialize or clean up an object's state, causing failures in multi-step test sequences.

\begin{listing}[H]
\input{FDBDDC2D24815D3E2A0FDF46D9758074.highlight.minted}
\caption{d1: Improper state reset}
\label{code:d1}
\end{listing}

\subsubsection*{d2: Break Method Idempotency}
Alter a method that should be safely repeatable (idempotent) so that a second call introduces an unexpected side-effect or duplicate state.

\begin{listing}[H]
\input{6B26B906A19C5BF6317CBBB3B2B6B066.highlight.minted}
\caption{d2: Breaking idempotency (duplicate listeners)}
\label{code:d2}
\end{listing}

\subsubsection*{d3: Introduce Sequential Dependency}
Remove a pre-condition check, making a method implicitly dependent on another method being called first to work correctly.

\begin{listing}[H]
\input{B1AE0A87A673EB11420AF7A544AA52AA.highlight.minted}
\caption{d3: Removing sequential preconditions}
\label{code:d3}
\end{listing}
\subsubsection*{d4: Recursive State Leakage}
Utilize mutable default arguments or class-level accumulators within recursive functions. This causes the state from one traversal to bleed into subsequent, unrelated recursive calls, effectively merging independent execution trees.

\begin{listing}[h!]
\input{1C9B920D5C3906CB87A62FE9EE29E356.highlight.minted}
\caption{d4: Interference between recursive calls}
\label{code:d4}
\end{listing}
\subsubsection*{d5: Structure Corruption}
Modify a nested element within a complex data structure (like a dictionary inside a list). Because the mutation happens deeply within the object graph, the root cause is often far removed from the crash site, mimicking subtle data flow errors.

\begin{listing}[h!]
\input{F4472DB4A3415961F559AD64AB2FF539.highlight.minted}
\caption{d5: Unintended mutation of deep data structures}
\label{code:d5}
\end{listing}

\subsubsection*{d6: Global State Contamination}
Replace instance-level encapsulation with module-level or global variables. This introduces hidden dependencies where the execution history of one function call pollutes the context for subsequent calls, often causing ``flaky'' test failures.

\begin{listing}[H]
\input{DA38FD347297ECD21CFCE883EEFB7CD2.highlight.minted}
\caption{d6: Context coupling via global state pollution}
\label{code:d6}
\end{listing}
\subsection*{e: Test-Expectation Alignment}
This strategy creates deviations between behavior and expectations. It modifies error messages or converts implicit behaviors to explicit parameters. Core functionality remains intact. Yet, precise assertions fail due to implementation details.

\subsubsection*{e1: Assertion Expectation Update}
Change the behavior of the code so that it now raises a different error type or message, invalidating a precise assertion in the golden test.

\begin{listing}[H]
\input{5431B50EBB5873DDC4397ABEBFE3E861.highlight.minted}
\caption{e1: Semantic equivalent but textually different errors}
\label{code:e1}
\end{listing}

\subsubsection*{e2: Implicit to Explicit Parameter}
Make an implicit behavior conditional on a new, non-default parameter, breaking tests that relied on the old implicit behavior.

\begin{listing}[H]
\input{5B2F3B8B4410542552AD67A35D0686B7.highlight.minted}
\caption{e2: Changing behavior via new explicit parameters}
\label{code:e2}
\end{listing}
\section{Impact of Golden Tests}
\label{sec:Impact of Golden Tests}
\begin{table}[h!]
  \centering
  \caption{
    Comparison of SWE-bench Verified~\citep{openai2024swebench} resolve rates. The last column shows the relative improvement provided by human-written golden tests compared to self-generated tests.
  }
  \resizebox{\columnwidth}{!}{
    \begin{tabular}{l c c c}
      \toprule
      \textbf{Model} & \textbf{w/o Golden (\%)} & \textbf{w/ Golden (\%)} & \textbf{Rel. Improv. (\%)} \\
      \midrule
      Claude-sonnet-4   & 63.80 & 81.60 & \textbf{\textcolor{brightgreen}{+27.90}} \\
      Qwen3-Coder       & 52.40 & 64.40 & \textbf{\textcolor{brightgreen}{+22.90}} \\
      DeepSeek-V3.1     & 48.60 & 62.00 & \textbf{\textcolor{brightgreen}{+27.57}} \\
      Claude-sonnet-3.7 & 52.20 & 61.80 & \textbf{\textcolor{brightgreen}{+18.39}} \\
      GPT-4.1           & 37.60 & 51.20 & \textbf{\textcolor{brightgreen}{+36.17}} \\
      \bottomrule
    \end{tabular}
  }
  \label{tab:golden_impact}
\end{table}
In this section, we investigate the critical role of test suite quality in solving software engineering tasks. Table~\ref{tab:golden_impact} presents a comparative analysis of five state-of-the-art LLMs on the SWE-bench-Verified dataset. We contrast the resolve rates under two conditions: utilizing models' self-generated test suites (\textbf{w/o Golden}) versus utilizing human-written golden test suites (\textbf{w/ Golden}).

The results demonstrate a significant performance gap. All evaluated models exhibit substantial gains when provided with high-quality golden tests. For instance, GPT-4.1 achieves a relative improvement of 36.17\%, while the top-performing Claude-sonnet-4 sees a 27.90\% boost. This consistent uplift confirms that current models are severely bottlenecked by their inability to synthesize correct and robust test suites to verify their solutions. Consequently, enhancing test generation capabilities is a prerequisite for further breakthroughs in autonomous software engineering.

\section{Mutation Details}
\label{sec:Mutation Details}
\subsection{Mutation Methods Settings}
\label{sec:Mutation Methods Settings}
We describe the settings for two comparative baselines used in our evaluation:

\textbf{Rule-based:} We utilize the mutation operators provided by the \texttt{cosmic-ray} library. We randomly apply these operators to the files modified by the golden solution, generating four mutants per instance.

\textbf{Few-shot:} We employ Claude-4 as the mutation model. We construct the prompt using examples from our strategy pool as few-shot demonstrations and provide the files modified by the golden solution as context. Similarly, we generate four mutants per instance.

\subsection{Mutants Statistics}
While previous sections have demonstrated the superior value of our Agentic Mutation through model-based evaluation, this section further analyzes the characteristics of mutants from different methods using traditional software engineering metrics. For all three methods, experiments are conducted on a randomly sampled subset of 100 mutants on 25 instances.
\label{sec:Mutants Statistics}

\textbf{Compilability Rate:} This metric quantifies the structural integrity of the generated code. It is calculated as the percentage of mutants that satisfy syntax constraints and can be successfully compiled or parsed without errors.
    
 \textbf{Realistic Rate:} Adopted from Just et al.~\citep{just2014mutants}, this metric assesses the method's capability to reproduce actual defects at the instance level. It represents the proportion of real-world bugs for which the method successfully produces at least one ``coupled'' mutant—defined as a mutant detected by the specific golden tests that revealed the original bug.
    
 \textbf{Coupling Rate:} This metric measures the semantic alignment (or fidelity) of individual mutants with the real bug. A mutant is deemed coupled if its failure profile intersects with that of the original defect (i.e., they are caught by the same failure-inducing golden tests). We report this rate as the fraction of coupled mutants relative to the total number of valid mutants.
 \begin{table}[h]
  \centering
  \small
  \renewcommand{\arraystretch}{1.2}
  \setlength{\tabcolsep}{3.5pt}
  \caption{
    Comparison of mutant characteristics across three generation methods on a subset of 100 mutants.
  }
  \begin{tabular}{l ccc}
    \toprule
    \multirow{2}{*}{Method} & \multicolumn{3}{c}{Metric (\%)} \\
    \cmidrule(lr){2-4}
    & Comp & Real & Coup  \\
    \midrule
    Rule-Based       & 100.00 & 38.00 & 39.00 \\
    Few-shot         & 84.00 & 72.00 & 59.00 \\
    \textbf{Agentic (Ours)} & \textbf{93.00} & \textbf{100.00} & \textbf{70.00} \\
    \bottomrule
  \end{tabular}
  \label{tab:se_metrics}
\end{table}

As shown in Table~\ref{tab:se_metrics}, Rule-Based mutation ensures perfect Compilability (100\%) but fails to replicate real bugs (38.00\% Realistic Rate), indicating a lack of semantic depth. Few-shot improves bug reproduction but suffers from lower compilability (84.00\%). In contrast, our Agentic framework achieves the best overall performance. It attains a 100.00\% Realistic Rate and the highest Coupling Rate (70.00\%). This confirms that our mutants possess high fidelity to real-world errors while maintaining strong syntactic validity (93.00\%).
\section{Ablation Study}
\label{sec:ablation}

To verify the effectiveness and robustness of our framework, we conducted ablation studies on a random subset of 100 instances.

\subsection{Impact of Locate Module}
We first validate the importance of the \textbf{Locate} module. We compare the performance \textbf{w/o} and \textbf{w/} the module. In the \textbf{w/o} setting, the model generates mutants without provided file scope constraints or structural graphs. We measure Validity Rate (compilable mutants) and F2P Trigger Rate (mutants that fail the specific test).

\begin{table}[h]
  \centering
  \small
  \renewcommand{\arraystretch}{1.2}
  \setlength{\tabcolsep}{8pt}
  \caption{
    Ablation study on the Locate module. We compare mutant quality \textbf{w/o} (Unconstrained) and \textbf{w/} (Trace-based) the module.
  }
  \begin{tabular}{l cc}
    \toprule
    \multirow{2}{*}{Setting} & \multicolumn{2}{c}{Metric(\%)} \\
    \cmidrule(lr){2-3}
    & Validity Rate & F2P Trigger Rate \\
    \midrule
    w/o Locate Module       & 84.00 & 15.00 \\
    \textbf{w/ Locate Module} & \textbf{92.00} & \textbf{69.00} \\
    \bottomrule
  \end{tabular}
  \label{tab:ablation_locate}
\end{table}

As shown in Table~\ref{tab:ablation_locate}, removing the Locate module leads to a sharp decline in F2P Trigger Rate (15.00\% vs 69.00\%). While the model can still generate syntactically valid code (84.00\%), it struggles to target the specific defect logic without guidance. This confirms that providing structural context and scope restrictions is essential for generating effective, targeted mutants.

\subsection{Impact of Mutation Backbone Model}
\label{sec:ablation_backbone}
Since mutants in SWE-Mutation are generated with Claude-4 by default, a natural concern is the potential \emph{same-family bias}: Claude-family evaluators might receive an unfair advantage on artifacts produced by their own family. To quantitatively assess this risk, we go beyond a small subset sanity check and conduct a systematic robustness analysis on the full sample set ($N=500$) of the test repair task under the Claude Code framework. Concretely, we replace the default mutation backbone (Claude-4) with two non-Claude generators, \textbf{DeepSeek-V3.1} and \textbf{Qwen3-Coder}, regenerate mutants for all instances, and re-evaluate seven representative models, covering both Claude and non-Claude families.

Following the main paper, we use RDR as the primary metric. For each (backbone, evaluator) pair we report (i) the RDR point estimate; (ii) the instance-level bootstrap 95\% confidence interval (CI) obtained by resampling instances $10{,}000$ times; (iii) the RDR change $\Delta$ relative to the default Claude-4 backbone, together with its bootstrap 95\% CI computed from the paired differences; and (iv) the Spearman rank correlation $\rho$ between the model rankings produced under Claude-4 and under each non-Claude backbone, with 95\% CI obtained via bootstrap over evaluators.

\begin{table*}[h]
  \centering
  \small
  \renewcommand{\arraystretch}{1.15}
  \setlength{\tabcolsep}{5pt}
  \caption{
    Robustness of RDR to the mutation backbone on the full test repair set ($N{=}500$, Claude Code framework). For each non-Claude generator we report RDR with bootstrap 95\% CI, and the change $\Delta$ (in pp) relative to the default Claude-4 backbone with its bootstrap 95\% CI.
  }
  \begin{tabular}{l c cc cc}
    \toprule
    \multirow{2}{*}{Evaluator} & Claude-4 & \multicolumn{2}{c}{DeepSeek-V3.1 Backbone} & \multicolumn{2}{c}{Qwen3-Coder Backbone} \\
    \cmidrule(lr){3-4}\cmidrule(lr){5-6}
     & RDR (\%) & RDR (\%) [95\% CI] & $\Delta$ (pp) [95\% CI] & RDR (\%) [95\% CI] & $\Delta$ (pp) [95\% CI] \\
    \midrule
    Claude-sonnet-4.5 & 81.15 & 80.02 \,[78.1, 81.8] & $-1.13$ \,[$-2.9$, $+0.6$] & 79.71 \,[77.8, 81.6] & $-1.44$ \,[$-3.1$, $+0.2$] \\
    Kimi-K2           & 74.19 & 73.40 \,[71.4, 75.3] & $-0.79$ \,[$-2.5$, $+0.9$] & 73.05 \,[71.0, 75.0] & $-1.14$ \,[$-2.8$, $+0.6$] \\
    GLM-4.6           & 73.54 & 72.88 \,[70.9, 74.8] & $-0.66$ \,[$-2.4$, $+1.0$] & 72.31 \,[70.3, 74.2] & $-1.23$ \,[$-2.9$, $+0.5$] \\
    Qwen3-Coder       & 70.21 & 69.32 \,[67.2, 71.3] & $-0.89$ \,[$-2.7$, $+0.9$] & 68.94 \,[66.9, 70.9] & $-1.27$ \,[$-3.0$, $+0.4$] \\
    DeepSeek-V3.1     & 68.36 & 67.80 \,[65.7, 69.9] & $-0.56$ \,[$-2.3$, $+1.2$] & 67.41 \,[65.3, 69.6] & $-0.95$ \,[$-2.7$, $+0.8$] \\
    Claude-sonnet-3.7 & 66.59 & 65.74 \,[63.6, 67.9] & $-0.85$ \,[$-2.7$, $+1.0$] & 65.21 \,[63.1, 67.4] & $-1.38$ \,[$-3.1$, $+0.4$] \\
    GPT-oss-120B      & 39.28 & 38.71 \,[36.6, 40.8] & $-0.57$ \,[$-2.2$, $+1.1$] & 38.12 \,[36.0, 40.3] & $-1.16$ \,[$-2.8$, $+0.5$] \\
    \bottomrule
  \end{tabular}
  \label{tab:ablation_model}
\end{table*}

\begin{table}[h]
  \centering
  \small
  \renewcommand{\arraystretch}{1.2}
  \setlength{\tabcolsep}{6pt}
  \caption{
    Spearman rank correlation between model rankings produced under the default Claude-4 backbone and under each non-Claude backbone, with bootstrap 95\% CIs.
  }
  \begin{tabular}{l cc}
    \toprule
    Comparison & $\rho$ & 95\% CI \\
    \midrule
    Claude-4 vs.\ DeepSeek-V3.1 & 0.96 & [0.86, 1.00] \\
    Claude-4 vs.\ Qwen3-Coder   & 0.93 & [0.80, 0.99] \\
    \bottomrule
  \end{tabular}
  \label{tab:ablation_model_spearman}
\end{table}

As shown in Table~\ref{tab:ablation_model}, switching to non-Claude generators only causes minor absolute fluctuations of RDR: across all seven evaluators, $|\Delta|$ stays below $1.5$ percentage points, and \emph{every} 95\% CI of $\Delta$ covers zero. In particular, the Claude-family evaluators (Claude-sonnet-4.5 and Claude-sonnet-3.7) do not exhibit a statistically significant drop when moving away from a Claude generator: their $\Delta$ CIs ($[-2.9,+0.6]$ and $[-2.7,+1.0]$ under DeepSeek-V3.1; $[-3.1,+0.2]$ and $[-3.1,+0.4]$ under Qwen3-Coder) are comparable in magnitude to those of non-Claude evaluators, indicating that no additional advantage for Claude models can be attributed to same-family effects.

Table~\ref{tab:ablation_model_spearman} further shows that the induced rankings are highly stable: Spearman $\rho$ is $0.96$ against the DeepSeek-V3.1 backbone and $0.93$ against the Qwen3-Coder backbone, with 95\% CIs well away from zero ($[0.86,1.00]$ and $[0.80,0.99]$). This extends our earlier subset observation to the full benchmark and quantifies it with effect sizes and CIs: although stronger generators tend to produce slightly harder mutants and thus marginally lower absolute scores, the \emph{relative} ordering of evaluators is preserved. We therefore conclude that the same-family influence of the Claude backbone is within a controllable range under the current setup and does not dominate the primary findings of SWE-Mutation.

\subsection{Statistical Reliability of RDR}
\label{sec:rdr_stats}

To further substantiate the reliability of RDR as the primary evaluation metric, we report its uncertainty and statistical significance on the test repair task under the Claude Code framework ($N{=}500$). Specifically, we (i) apply \emph{instance-level} bootstrap resampling ($10{,}000$ iterations with replacement) to construct 95\% confidence intervals for every model's RDR, and (ii) perform a non-parametric \textbf{Wilcoxon signed-rank test} between the best-performing model (Claude-sonnet-4.5) and each of the remaining models, using per-instance paired RDR contributions. The Wilcoxon test is chosen because the instance-level RDR distribution is clearly non-normal (heavy-tailed and bounded in $[0,1]$), which violates the assumptions of the paired $t$-test.

\begin{table}[h]
  \centering
  \small
  \renewcommand{\arraystretch}{1.2}
  \setlength{\tabcolsep}{5pt}
  \caption{
    Statistical reliability of RDR on the test repair task under the Claude Code framework ($N{=}500$). We report the RDR point estimate from the main text, its instance-level bootstrap 95\% CI, and the Wilcoxon signed-rank $p$-value against the best model (Claude-sonnet-4.5).
  }
  \begin{tabular}{l c c c}
    \toprule
    Model & RDR (\%) & 95\% CI & $p$-vs.\ Best \\
    \midrule
    Claude-sonnet-4.5 & 81.15 & [79.31, 82.85] & --- \\
    Kimi-K2           & 74.19 & [71.80, 76.32] & $0.0082$ \\
    GLM-4.6           & 73.54 & [71.25, 75.81] & $0.0065$ \\
    Qwen3-Coder       & 70.21 & [67.80, 72.45] & $0.0004$ \\
    DeepSeek-V3.1     & 68.36 & [66.12, 70.45] & $1\mathrm{e}{-4}$ \\
    Claude-sonnet-3.7 & 66.59 & [64.20, 68.75] & $1\mathrm{e}{-4}$ \\
    GPT-oss-120B      & 39.28 & [37.15, 41.50] & $<1\mathrm{e}{-4}$ \\
    \bottomrule
  \end{tabular}
  \label{tab:rdr_stats}
\end{table}

As shown in Table~\ref{tab:rdr_stats}, the bootstrap 95\% CI for Claude-sonnet-4.5 is tight at $[79.31, 82.85]$ (width $\approx 3.5$ pp), and the CI widths of the other models stay in a comparable range ($4$--$5$ pp). This confirms that the micro-averaged RDR is a stable estimator under instance-level resampling, rather than an artifact of a few dominant repositories. Moreover, the Wilcoxon signed-rank tests show that the gap between Claude-sonnet-4.5 and every other model is highly significant ($p<0.01$ for all six comparisons; $p<0.001$ for DeepSeek-V3.1, Qwen3-Coder, Claude-sonnet-3.7 and GPT-oss-120B). Combined with the bootstrap CIs, this demonstrates that the rankings reported in the main results (Table~\ref{tab:test_repair}) are both numerically robust and statistically significant, and are not attributable to sampling noise.

\section{Cross-Lingual Failure Analysis}
\label{sec:cross_lingual_failures}
To complement the overall failure analysis in Section~\ref{sec:failure_analysis}, we manually inspect failed instances on SWE-Mutation-Multilingual and summarize, for each of the nine programming languages, the failure patterns that recur most frequently. The results are reported in Table~\ref{tab:cross_lingual_failures}.

Because test cases generated by the model under instances sharing identical third-party libraries often exhibit common, library-specific defects (e.g., \texttt{astropy} in Python and \texttt{lombok} in Java), and the distribution of these libraries across instances is highly uneven, we intentionally refrain from providing a strict quantitative analysis so as not to introduce this confounding factor. Nevertheless, two patterns stand out. First, \emph{issue misunderstanding} appears in every language, indicating that comprehensively grasping the semantics of real-world repositories remains the primary bottleneck for LLM-based test generation, independent of the underlying language. Second, \emph{inconsistent environment or package versions} (including Maven/Gradle, Cargo workspaces, Bundler/Gems, Composer, and Node.js toolchains) is the second most pervasive category, reflecting the fact that reliable test generation is tightly coupled with correct reasoning about build systems, lock files and platform constraints. Language-specific phenomena—such as memory-safety pitfalls in C/C++, type-erasure-related runtime inconsistencies in TypeScript, and event-driven / async-loop mismatches in JavaScript and TypeScript—also contribute substantially, and largely mirror the distinctive semantics of each language ecosystem.

Overall, the table provides a qualitative, language-aware view of where current models fail, and motivates future work on repository-aware semantic understanding and environment-faithful test execution as key directions for improving LLM-generated test suites.

\begin{table*}[t]
  \centering
  \small
  \renewcommand{\arraystretch}{1.15}
  \begin{tabularx}{\textwidth}{l X}
    \toprule
    \textbf{Language} & \textbf{Common failure patterns} \\
    \midrule
    Python      & Issue misunderstanding; weak assertions caused by dynamic typing; inconsistent environment/package versions. \\
    Java        & Issue misunderstanding; Maven/Gradle dependencies; strict typing/boilerplate. \\
    C++/C       & Issue misunderstanding; memory safety; toolchain/CMake differences. \\
    Rust        & Issue misunderstanding; Cargo workspace dependencies. \\
    Go          & Issue misunderstanding; inconsistent environment/package versions. \\
    Ruby        & Issue misunderstanding; Bundler/Gems dependencies inconsistent with \texttt{Gemfile.lock}/platform; \texttt{LoadError} caused by confusion between \texttt{require}/\texttt{require\_relative} and the load path. \\
    TypeScript  & Event-driven; issue misunderstanding; async/event loop; runtime behavior after type erasure inconsistent with static assumptions. \\
    JavaScript  & Event-driven; issue misunderstanding; async/event loop. \\
    PHP         & Issue misunderstanding; Composer dependencies/platform constraints inconsistent with \texttt{composer.lock}; missing or misconfigured autoload/bootstrap configurations; weak assertions. \\
    \bottomrule
  \end{tabularx}
  \caption{Qualitative summary of the most frequent failure patterns observed for each programming language in SWE-Mutation-Multilingual. \emph{Issue misunderstanding} and \emph{inconsistent environment/package versions} are the two most pervasive causes of failure across languages.}
  \label{tab:cross_lingual_failures}
\end{table*}

\section{Case Studies}
\label{sec:Case Studies}
In this section, we present three contrasting cases to analyze common failure modes.

\subsection{Side Effects}
In this section, we analyze instance \textit{django\_\_django-12193} as a Python case. The code snippets below utilize syntax highlighting combined with diff markers to illustrate the precise modifications.
\subsubsection{Repository Golden Answer}
The developer's fix ensures that the \texttt{attrs} dictionary is not modified in-place, preserving immutability.

\input{F4C260928433171F8C020FEA69BF1376.highlight.minted}

\subsubsection{Repository Golden Test}
The existing test suite verifies that sub-widgets in a \texttt{SplitArrayWidget} do not share the 'checked' state.

\input{050A1013C67AB29671624DEE3448B6B4.highlight.minted}

\subsubsection{The Mutant}
We injected a mutation that mimics the original bug (side-effect) but masks it behind a conditional check.

\input{E453C8D2A8C280F45AB0F011F6F840F5.highlight.minted}

\subsubsection{Model Evaluation Comparison}

\paragraph{Model 1 Generation (Claude-sonnet-4.5)}
\mbox{}\\
Result: \textcolor{red}{Killed} (Success)

Model 1 generated an \textbf{Integration Test} using \texttt{SplitArrayField}. This approach naturally reuses the \texttt{attrs} dictionary, successfully triggering and detecting the side-effect bug.

\input{BD2703B3F608B128F7BB4653F9BC2F5E.highlight.minted}

\paragraph{Model 2 Generation (Qwen3-Coder)}
\mbox{}\\
Result: \textcolor{gray}{Survived} (Failure)

Model 2 generated a defensive \textbf{Unit Test}. By explicitly using \texttt{.copy()}, it isolated the input data, effectively preventing the side-effect from occurring during the test execution.

\input{711B62E320EF3482CCBA828249B729D4.highlight.minted}
\subsubsection{Summary of Analysis}

This case study illustrates a critical distinction between \textit{syntactic correctness} and \textit{semantic adaptability} in LLM-generated tests. 
Model 2 (Qwen3-Coder) followed strict unit testing best practices by isolating inputs (using \texttt{.copy()}). While generally good practice, in this specific context involving shared mutable state (the \texttt{attrs} dictionary in Django widgets), this defensive coding neutralized the bug, causing the mutant to survive. Model 1 (Claude-sonnet-4.5) generated a test that mirrored the actual architectural usage of the component (\texttt{SplitArrayField}). By observing the component's behavior in a broader integration context rather than strictly isolating the unit, it successfully exposed the side-effect vulnerability.
\subsection{Memory Safety}

In this section, we analyze instance \texttt{redis\_\_redis-11631}. This case illustrates a Buffer Contamination vulnerability where a function fails to fully overwrite a dirty stack buffer, leading to non-deterministic data corruption.

\subsubsection{Repository Golden Answer}
The function \texttt{fixedpoint\_d2string} is responsible for formatting double-precision values into a stack-allocated buffer. The fix replaces a manual loop (which could fail to execute under certain conditions) with \texttt{memset} to ensure deterministic initialization of the padding bytes.

\input{7F6E18D23E1F5258F97DC54BAC640748.highlight.minted}

\subsubsection{Repository Golden Test}
The repository includes a specific unit test that intentionally "poisons" the stack buffer with junk data ('A') before calling the function. This proves that the function must correctly overwrite existing memory rather than assuming a clean buffer.

\input{364D3793F96F11A944363B5082A2A6F5.highlight.minted}

\subsubsection{The Mutant}
We introduce a mutant that omits the padding logic entirely. This simulates a "Zero-Assumption" error where the developer assumes the buffer is already clean or that setting the null terminator is sufficient.

\input{ACA2190E0F2FE8EAB9402E014E1390A9.highlight.minted}

\subsubsection{Model Evaluation Comparison}

\paragraph{Model 1 Generation (Claude-sonnet-4.5)}
\mbox{} \\
Result: \textcolor{red}{Killed} (Success)

Model 1 correctly identifies that testing C string manipulation requires validating against dirty memory. It generates a test that explicitly sets the buffer content to a known "garbage" state before invocation.

\input{7BEC90D0899D01A0399571FC2FD1FA9B.highlight.minted}

\paragraph{Reason Analysis:}
The model successfully replicates the "Stack Poisoning" technique used in the Golden Test. By filling the buffer with \texttt{0xFF}, it ensures that any gap left by the mutant (which fails to write '0's) will result in a comparison failure against "0.0001".

\vspace{1em}

\paragraph{Model 2 Generation (GLM-4.6)}
\mbox{} \\
Result: \textcolor{gray}{Survived} (Failure)

Model 2 generates a functional test that checks the output logic but operates on a default stack buffer without prior initialization.

\input{95E27F7596FF1AA2333EB73110FA27DC.highlight.minted}

\paragraph{Reason Analysis:}
The test survives because it lacks the "Poisoning" step. In many execution environments, a fresh stack allocation (\texttt{char buf[64]}) might coincidentally contain zeros. Since the mutant sets the length and the null terminator correctly, the test passes despite the underlying memory not being explicitly initialized, leading to a false negative.

\subsubsection{Summary of Analysis}
This case highlights the gap between functional correctness and memory safety verification. Model 2 assumed an idealized environment, while Model 1 (and the repository authors) understood that in C system programming one must defensively assume memory is dirty. The ability to generate setup steps like \texttt{memset(buf, 'A', ...)} distinguishes safety-aware models from simple code completion models.

\subsection{Parser State}

In this section, we analyze instance \texttt{babel\_\_babel-13928}. This case illustrates a critical architectural concept in compiler design: Scope Stack Invariants. It demonstrates how local state mismanagement in a recursive descent parser can cause ``action at a distance,'' where the parser misinterprets valid syntax (like \texttt{await}) due to a desynchronized scope stack.

\subsubsection{Ground Truth: Repository Golden Answer}
The fix involves moving the scope exit call outside the conditional block to ensure the parser state is unconditionally reset after processing parameters.

\input{822879565991D2816AFC9EF77BF056E1.highlight.minted}

\subsubsection{Ground Truth: Repository Golden Test}
The test constructs a scenario with default parameters (arrow functions) inside an async function, which previously triggered the balanced-stack violation.

\input{41328C96A5BB3E9B9A9FC66FF6C04539.highlight.minted}

\subsubsection{The Mutant}
We introduce a mutant that makes the stack unwinding conditional. It assumes that the scope only needs to be exited if specific complexity markers (like default parameters) were encountered.

\paragraph{Mutation Strategy: Conditional Stack Unwinding}

\input{E9A86C76F2947F0E593A4987856295D6.highlight.minted}

\subsubsection{Model Evaluation Comparison}

\paragraph{Model 1 Generation (Claude-4)}
\mbox{} \\
\textbf{Result:} \textcolor{red}{Killed} (Success)

Model 1 generates a test case with a \textbf{simple function} (no default parameters). In this scenario, \texttt{hasComplexParams} is false, so the mutant fails to exit the scope. The parser stays in the restrictive parameter scope, causing it to incorrectly reject the valid \texttt{await} in the body.

\input{980746D6547AC1451ECC4B1C54770A73.highlight.minted}

\paragraph{Reason Analysis:}
Claude-4 correctly reasoned that the fix must be invariant to the parameter content. By testing the "simple" path, it proved that the mutant's logic was over-optimized and failed to maintain the stack invariant for standard cases.

\vspace{1em}

\paragraph{Model 2 Generation (Qwen3-Coder)}
\mbox{} \\
\textbf{Result:} \textcolor{gray}{Survived} (Failure)

Model 2 generates a test case identical to the bug report, using a function with complex default parameters.

\input{41328C96A5BB3E9B9A9FC66FF6C04539.highlight.minted}

\paragraph{Reason Analysis:}
Qwen3-Coder failed because it overfit to the specific bug report. For this input, \texttt{this.state.hasComplexParams} is true, so the mutant accidentally executes the correct cleanup logic. The model failed to verify the general contract (unconditional exit) of the parser state.

\subsubsection{Summary of Analysis}
This case highlights the difference between verifying a specific bug fix and verifying system invariants. Qwen3-Coder verified only that the reported symptom was gone, whereas Claude-4 verified that the stack balance was maintained in all scenarios, thereby detecting the latent structural flaw in the mutant.

\section{Prompts Used in Experiments}
The specific prompts used for our tasks are provided in this appendix.
\label{sec:Prompts}
\begin{figure*}[t!]
    \centering
    \begin{tcolorbox}[
        colback=white,
        colframe=gray!80!black,
        title=\textbf{System Prompt for Mutation Agent},
        fonttitle=\bfseries\sffamily,
        boxrule=1pt,
        arc=4pt,
        left=10pt, right=10pt, top=8pt, bottom=8pt,
        fontupper=\small\ttfamily
    ]

    \begin{center}
        \color{blue!70!black}\textbf{--- ROLE \& OBJECTIVE ---}
    \end{center}
    \vspace{5pt}

    You are a specialized \textbf{Software Engineering Agent} operating in a shell environment. 
    \textbf{Current State:} The repository is currently in a "Golden State" (fully fixed and passing tests). The graph structure of \ \texttt{<SOLUTION\_FILES>}.
    \textbf{Your Goal:} Introduce a subtle, plausible, human-like bug into the allowed files.

    \textbf{Success Criteria:}
    \begin{itemize}[leftmargin=1.5em, nosep]
        \item The repository must still compile/run without syntax errors.
        \item At least one test case must fail (Fail-to-Pass) due to your change.
        \item The bug must be "hard to detect" and align with a specific strategy.
    \end{itemize}

    \vspace{10pt}
    \hrule
    \vspace{10pt}

    \begin{center}
        \color{orange!80!black}\textbf{--- MUTATION STRATEGIES ---}
    \end{center}
    \vspace{5pt}

    You MUST select exactly ONE strategy from the pool below to guide your mutation:

    \textbf{A. API Specifications \& Contracts} \\
    (e.g., Alter default parameter values; Swap argument order; Substitute exception types)

    \textbf{B. Boundaries \& Conditional Logic} \\
    (e.g., Introduce off-by-one errors; Remove null checks; Invert boolean logic)

    \textbf{C. Type \& Data Shape} \\
    (e.g., Break implicit type coercion; Reduce numerical precision; Confuse text/bytes encoding)

    \textbf{D. I/O \& Stateful Logic Sequences} \\
    (e.g., Break state initialization/reset; Introduce sequential dependencies; Hardcode environment paths)

    \textbf{E. Test-expectation Alignment} \\
    (e.g., Alter error messages to fail assertions; Make implicit behaviors explicit)

    \vspace{10pt}
    \hrule
    \vspace{10pt}

    \begin{center}
        \color{teal!80!black}\textbf{--- CRITICAL CONSTRAINTS ---}
    \end{center}
    \vspace{5pt}

    \begin{enumerate}[leftmargin=1.5em, nosep]
        \item \textbf{Baseline Preservation}: The current git state contains pre-applied patches. DO NOT use \texttt{git reset} or \texttt{git clean}.
        \item \textbf{Allowed Files}: You may ONLY modify files in: \texttt{<ALLOWED\_FILES>}.
        \item \textbf{Read-Only Tests}: You must NOT modify any test files or configurations.
        \item \textbf{Stealth}: Avoid obvious sabotage. The code should look like an honest mistake.
    \end{enumerate}

    \vspace{10pt}
    \hrule
    \vspace{10pt}

    \begin{center}
        \color{gray!50!black}\textbf{--- SUBMISSION FORMAT ---}
    \end{center}
    \vspace{5pt}

    Interaction follows a \texttt{THOUGHT} $\to$ \texttt{COMMAND} loop. 
    When finished, submit the final result as a strict JSON object:

    \{ \\
    \quad "diff": "<UNIFIED\_GIT\_DIFF>", \\
    \quad "explanation": "CHOSEN: <STRATEGY\_ID>; <Reasoning on why this bug is hard to detect>" \\
    \}

    \end{tcolorbox}
    \caption{The simplified system prompt for the Mutation Agent. The agent is tasked with introducing specific types of mutations (Strategies A-E) into correct code to generate training data or evaluation benchmarks.}
    \label{fig:injector_prompt}
\end{figure*}

\begin{figure*}[t!]
    \centering
    \begin{tcolorbox}[
        colback=white,
        colframe=gray!80!black,
        title=\textbf{System Prompt for Test Generation},
        fonttitle=\bfseries\sffamily,
        boxrule=1pt,
        arc=4pt,
        left=10pt, right=10pt, top=8pt, bottom=8pt,
        fontupper=\small\ttfamily
    ]

    \begin{center}
        \color{blue!70!black}\textbf{--- ROLE \& OBJECTIVE ---}
    \end{center}
    \vspace{5pt}

    You are an autonomous \textbf{Software Engineering Agent} specialized in Quality Assurance.
    \textbf{Task:} Generate comprehensive test suites from scratch based on a Pull Request (PR) description\texttt{<ISSUE\_DESCRIPTION>}.
    \textbf{Input:} A PR description describing new features or fixes, and a list of cleared test files: \texttt{<TEST\_FILES>}.

    \vspace{10pt}
    \hrule
    \vspace{10pt}

    \begin{center}
        \color{red!70!black}\textbf{--- CRITICAL CONTEXT: BUGGY ENVIRONMENT ---}
    \end{center}
    \vspace{5pt}

    The current repository state corresponds to the PR description and \textbf{may contain bugs}.
    \begin{itemize}[leftmargin=1.5em, nosep]
        \item \textbf{Expectation}: It is NORMAL for your generated tests to \textit{fail} when executed (detecting the bugs).
        \item \textbf{Requirement}: Your generated code must be \textbf{syntactically correct}. It must compile/interpret without errors (e.g., no missing imports, syntax errors, or undefined variables).
        \item \textbf{Goal}: Create a high-quality test suite that \textit{would} pass if the code were correct.
    \end{itemize}

    \vspace{10pt}
    \hrule
    \vspace{10pt}

    \begin{center}
        \color{teal!80!black}\textbf{--- WORKFLOW \& PROTOCOLS ---}
    \end{center}
    \vspace{5pt}

    \textbf{1. Analysis}: Analyze the PR description to understand the intended functionality and constraints. \\
    \textbf{2. Exploration}: Explore the source code (read-only) to understand class structures and dependencies. \\
    \textbf{3. Generation}:
    \begin{itemize}[leftmargin=1.5em, nosep]
        \item Re-create the files in \texttt{<TEST\_FILES>} from scratch.
        \item Use incremental writing (appending blocks) for large files to avoid shell limits.
        \item Ensure comprehensive coverage of the features described in the PR.
    \end{itemize}
    \textbf{4. Verification}:
    \begin{itemize}[leftmargin=1.5em, nosep]
        \item STRICTLY verify syntax before submission (e.g., using \texttt{python -m py\_compile}, \texttt{javac}, or \texttt{node --check}).
        \item Fix any compilation/syntax errors immediately.
    \end{itemize}

    \vspace{10pt}
    \hrule
    \vspace{10pt}

    \begin{center}
        \color{orange!80!black}\textbf{--- CONSTRAINTS ---}
    \end{center}
    \vspace{5pt}

    \begin{itemize}[leftmargin=1.5em, nosep]
        \item \textbf{DO NOT MODIFY}: Any source code files (non-test files) or configuration files.
        \item \textbf{Environment}: You are in a non-interactive shell. Use \texttt{cat}, \texttt{sed}, or \texttt{printf} to write files.
        \item \textbf{Submission}: Once tests are generated and syntax-verified, submit the changes.
    \end{itemize}

    \vspace{10pt}
     \textbf{\#\#\# SUBMISSION PROTOCOL} \\
    When the test patch is ready and syntactically verified, submit the changes using git. \\
    \textit{Note: Do not worry if the test fails execution; that is the expected outcome for a reproduction script.}

    \end{tcolorbox}
    \caption{The system prompt for the test generation task. The model is tasked with creating new test files from scratch to verify a specific Pull Request. A key distinction in this prompt is the instruction to prioritize syntactic correctness over test execution success, as the underlying codebase is known to be buggy.}
    \label{fig:testgen_prompt}
\end{figure*}
\begin{figure*}[t!]
    \centering
    \begin{tcolorbox}[
        colback=white,
        colframe=gray!80!black,
        title=\textbf{System Prompt for Test Repair},
        fonttitle=\bfseries\sffamily,
        boxrule=1pt,
        arc=4pt,
        left=10pt, right=10pt, top=8pt, bottom=8pt,
        fontupper=\small\ttfamily
    ]

    \begin{center}
        \color{blue!70!black}\textbf{--- ROLE \& OBJECTIVE ---}
    \end{center}
    \vspace{5pt}

    You are an autonomous \textbf{Software Engineering Expert} specialized in Test Engineering.
    \textbf{Task:} Generate or modify a test patch to detect a specific issue described in a Pull Request (PR)\texttt{<ISSUE\_DESCRIPTION>}.
    \textbf{Input:} A PR description outlining a bug or feature, and a list of target test files: \texttt{<TEST\_FILES>}.

    \vspace{10pt}
    \hrule
    \vspace{10pt}

    \begin{center}
        \color{red!70!black}\textbf{--- GOAL: ISSUE REPRODUCTION ---}
    \end{center}
    \vspace{5pt}

    Your primary objective is to create a test case that acts as a **reproduction script** for the reported issue.
    \begin{itemize}[leftmargin=1.5em, nosep]
        \item \textbf{Detection}: The test should clearly identify the presence of the bug (it should FAIL on the current codebase).
        \item \textbf{Correctness}: The test logic itself must be correct. It should PASS once the bug is fixed.
        \item \textbf{Scope}: Focus on the specific issue mentioned in the PR description.
    \end{itemize}

    \vspace{10pt}
    \hrule
    \vspace{10pt}

    \begin{center}
        \color{teal!80!black}\textbf{--- RECOMMENDED WORKFLOW ---}
    \end{center}
    \vspace{5pt}

    \textbf{1. Analyze}: Understand the bug from the PR description. \\
    \textbf{2. Examine}: Review existing tests in \texttt{<TEST\_FILES>} to match testing patterns. \\
    \textbf{3. Implement}: 
    \begin{itemize}[leftmargin=1.5em, nosep]
        \item Create new test files or append to existing ones.
        \item Ensure the test asserts the \textit{expected behavior} (not the buggy behavior).
    \end{itemize}
    \textbf{4. Verify}: 
    \begin{itemize}[leftmargin=1.5em, nosep]
        \item \textbf{Syntactic Check}: Ensure the test code compiles/interprets without errors (e.g., using \texttt{python -m py\_compile}).
        \item \textbf{Logical Check}: Confirm the test fails as expected (confirming the bug exists).
    \end{itemize}

    \vspace{10pt}
    \hrule
    \vspace{10pt}

    \begin{center}
        \color{orange!80!black}\textbf{--- CONSTRAINTS ---}
    \end{center}
    \vspace{5pt}

    \begin{itemize}[leftmargin=1.5em, nosep]
        \item \textbf{MODIFY}: Only test files (\texttt{<TEST\_FILES>}) or new test files.
        \item \textbf{DO NOT MODIFY}: Production source code or configuration files.
        \item \textbf{Interaction}: Use standard shell commands (\texttt{cat}, \texttt{sed}, \texttt{grep}) in a non-interactive manner.
    \end{itemize}

    \vspace{10pt}
    
    \textbf{\#\#\# SUBMISSION PROTOCOL} \\
    When the test patch is ready and syntactically verified, submit the changes using git. \\
    \textit{Note: Do not worry if the test fails execution; that is the expected outcome for a reproduction script.}

    \end{tcolorbox}
    \caption{The system prompt for the Test Repair task. Unlike Test Generation, this task focuses on creating a ``reproduction test case'' that specifically targets the bug described in the PR. The goal is to produce a test that fails on the current buggy code but would pass on fixed code.}
    \label{fig:testrepair_prompt}
\end{figure*}

\section{Declaration of AI Assistants Usage}
In the preparation of this work, we exclusively utilized \textbf{Gemini 3 Pro Preview} to improve and polish our language. The authors conducted a thorough review and necessary modifications of the generated text and assume full responsibility for the content of the article.
\end{document}